\documentclass[aps,prd,twocolumn,showpacs,groupedaddress,preprintnumbers,superscriptaddress,amsmath,amssymb,floatfix,longbibliography]{revtex4-1} 

\def\nn{\nonumber}

\newcommand{\bea}{\begin{eqnarray}}
\newcommand{\eea}{\end{eqnarray}}

\usepackage{color}			
\usepackage{graphicx}
\usepackage{bm}

\usepackage{physics}
\usepackage{braket}
\usepackage{slashed}

\usepackage[colorlinks=true, filecolor=blue, citecolor=blue,urlcolor=blue]{hyperref}
\usepackage{url}

\begin{document}
	
\abovedisplayskip = 4pt
\belowdisplayskip = 4pt
\abovedisplayshortskip = 4pt
\belowdisplayshortskip= 4pt

\title{Probing millicharged particles with NA64$\mu$ and LDMX} 

\author{Sergei N.~Gninenko} 
\affiliation{Institute for Nuclear Research, 117312 Moscow, Russia}
\affiliation{Bogoliubov Laboratory of Theoretical Physics, JINR, 141980 Dubna, Russia} 
\affiliation{Millennium Institute for Subatomic Physics at
the High-Energy Frontier (SAPHIR) of ANID, \\
Fern\'andez Concha 700, Santiago, Chile}

\author{ N.~V.~Krasnikov}
\affiliation{Institute for Nuclear Research, 117312 Moscow, Russia}
\affiliation{Bogoliubov Laboratory of Theoretical Physics, JINR, 141980 Dubna, Russia} 

\author{Sergey Kuleshov}
\affiliation{Millennium Institute for Subatomic Physics at
the High-Energy Frontier (SAPHIR) of ANID, \\
Fern\'andez Concha 700, Santiago, Chile}
\affiliation{Center for Theoretical and Experimental Particle Physics,
Facultad de Ciencias Exactas, Universidad Andres Bello,
Fernandez Concha 700, Santiago, Chile}

\author{Valery~E.~Lyubovitskij}
\affiliation{Institut f\"ur Theoretische Physik, Universit\"at T\"ubingen, \\
Kepler Center for Astro and Particle Physics, \\ 
Auf der Morgenstelle 14, D-72076 T\"ubingen, Germany} 
\affiliation{Millennium Institute for Subatomic Physics at
the High-Energy Frontier (SAPHIR) of ANID, \\
Fern\'andez Concha 700, Santiago, Chile}

\author{P. Crivelli}
\affiliation{ETH Zürich, Institute for Particle Physics and Astrophysics, CH-8093 Zürich, Switzerland}

\author{D.~V.~Kirpichnikov}
\email[\textbf{e-mail}: ]{dmbrick@gmail.com}
\affiliation{Institute for Nuclear Research, 117312 Moscow, Russia}

\author{L.~Molina Bueno}
\affiliation{CSIC - Universitat de València, Instituto de Física Corpuscular (IFIC), E-46980 Paterna, Spain}

\author{Alexey~S.~Zhevlakov } 
\affiliation{Bogoliubov Laboratory of Theoretical Physics, JINR, 141980 Dubna, Russia} 
\affiliation{Matrosov Institute for System Dynamics and 
	Control Theory SB RAS, \\  Lermontov str., 134, 664033, Irkutsk, Russia } 

\author{H.~Sieber}
\affiliation{ETH Zürich, Institute for Particle Physics and Astrophysics, CH-8093 Zürich, Switzerland}

\author{I.~V.~Voronchikhin}
\affiliation{ Tomsk Polytechnic University, 634050 Tomsk, Russia}
\affiliation{Institute for Nuclear Research, 117312 Moscow, Russia}

\begin{abstract}
Millicharged particles  emerge as compelling candidates in numerous theoretically well-motivated 
extensions of the Standard Model. These hypothetical particles, characterized by an electric charge that is a 
small fraction of the elementary charge, have attracted significant attention in contemporary experimental 
physics. Their potential existence motivates dedicated search strategies across multiple experimental platforms, 
leveraging their distinctive electromagnetic interactions while evading conventional detection methods.
In the present paper we estimated the projected sensitivity of fixed-target experiments, specifically NA64$\mu$ 
and LDMX, to the parameter space of millicharged particles.  
For the NA64$\mu$ experiment, with an anticipated muon flux of $\mbox{MOT}\lesssim 10^{14}$, our analysis 
reveals a detectable  mass window of $10~\mbox{MeV} \lesssim m_\chi \lesssim 150~\mbox{MeV}$ and charge parameter range $10^{-4} \lesssim  \epsilon \lesssim 7\times 10^{-4}$. This sensitivity arises from the bremsstrahlung-like missing energy signature $\mu N \to \mu N \gamma^{*}( \to \chi \bar{\chi})$.
Furthermore, we evaluate the discovery potential of the LDMX facility, considering its projected electron beam 
statistics, $\mbox{EOT}\lesssim  2\times 10^{16}$, and energy, $E_{\rm e}\simeq 8~\mbox{GeV}$. Our results 
demonstrate that LDMX can probe heavier MCPs in the mass range $250~\mbox{MeV} \lesssim  m_\chi \lesssim  400 ~\mbox{MeV}$, with 
sensitivities reaching $10^{-3} \lesssim  \epsilon \lesssim  1.5 \times 10^{-3}$.
This parametric window can be accessible through the distinctive 
invisible decay channel $\rho \to \chi \bar{\chi}$, where 
$\rho$-meson photo-production $\gamma N \to N \rho$ plays a pivotal role. 
\end{abstract}

\maketitle

\section{Introduction}

In scenarios
beyond the Standard Model (SM), particles with an electric charge significantly smaller than that 
of  the electron ($Q_{\chi} = e \epsilon \ll e$) emerge naturally. Notably, millicharged particles (MCPs) serve as 
well-motivated  candidates for dark matter (DM) or its fraction, as evidenced by extensive phenomenological  
studies~\cite{Brahm:1989jh,Cline:2012is,Pospelov:2007mp,Feng:2009mn,Tulin:2012wi,Feldman:2007wj}.  As a result, 
experimental probes for such particles hold  substantial importance in contemporary 
particle  physics~\cite{Feng:2022inv,Harnik:2019zee}.

A minimal approach to incorporating millicharged particles  into a theoretical framework is to 
treat them as a low-energy effective limit of a more fundamental theory. In this scenario, a hidden 
(dark) photon field, denoted as $A_\mu'$, has kinetic mixing 
$ \frac{1}{2} \epsilon F_{\mu \nu} F_{\mu \nu}'$ with the Standard Model photon 
$A_{\mu}$\,\cite{Holdom:1985ag}. In order to get rid of the kinetic mixing one can perform a redefinition of the field basis, $A_\mu' \to A_\mu' + \epsilon A_\mu$. Consequently, a 
hidden-sector fermion $\chi$, which couples to the 
dark photon, may obtain an effective electric charge proportional to $e \epsilon$. 
The corresponding Lagrangian  for this model can be expressed as
\begin{equation}
\mathcal{L} \supset i \bar{\chi} \gamma^\mu \partial_\mu \chi - m_{\chi} \bar{\chi} \chi + e \epsilon A_\mu \bar{\chi} \gamma^\mu \chi,
\end{equation}
where $m_\chi$ is the mass of the millicharged fermion $\chi$.

The phenomenological constraints on millicharged particles  can be comprehensively 
mapped onto the $(\epsilon, m_\chi)$ parameter space. 
Current and projected experimental limits imply several distinct searches: collider 
probes~\cite{SENSEI:2023gie,Liu:2018jdi,Zhang:2019wnz,Liu:2019ogn,Liang:2019zkb,Bai:2021nai,milliQan:2021lne}, 
neutrino detectors~\cite{LSND:2001akn},
experiments with a fixed
  target~\cite{Arefyeva:2022eba,Chu:2020ysb,Gorbunov:2021jog,ArgoNeuT:2019ckq}, proton beam dump~\cite{Marocco:2020dqu,Forbes:2024zks,Demidov:2025yyr,Bailloeul:2025fde,Allison:2025mom}, laboratory based tests~\cite{Dmitrieva:2025ohn,Berlin:2025btf,Berlin:2025hjs}, and 
  electron beam dump facility~\cite{Prinz:1998ua,Chu:2018qrm,Essig:2024dpa,Eberl:2025kfm}. Complementary constraints arise 
  from cosmological  observations and  astrophysical 
data~\cite{Harnik:2020ugb,Aboubrahim:2021ohe,Li:2020wyl,Caputo:2019tms,Kouvaris:2025tom}. Additional exclusion limits come from cosmic-ray 
detectors~\cite{ArguellesDelgado:2021lek,Plestid:2020kdm} and nuclear reactor 
experiments~\cite{CONNIE:2024off,TEXONO:2018nir,Chen:2014dsa,Gao:2025ykc}. These multi-pronged experimental efforts have
systematically excluded significant portions of the MCPs parameter space.

\begin{figure}[!tb]
\centering
\includegraphics[width=0.49\textwidth]{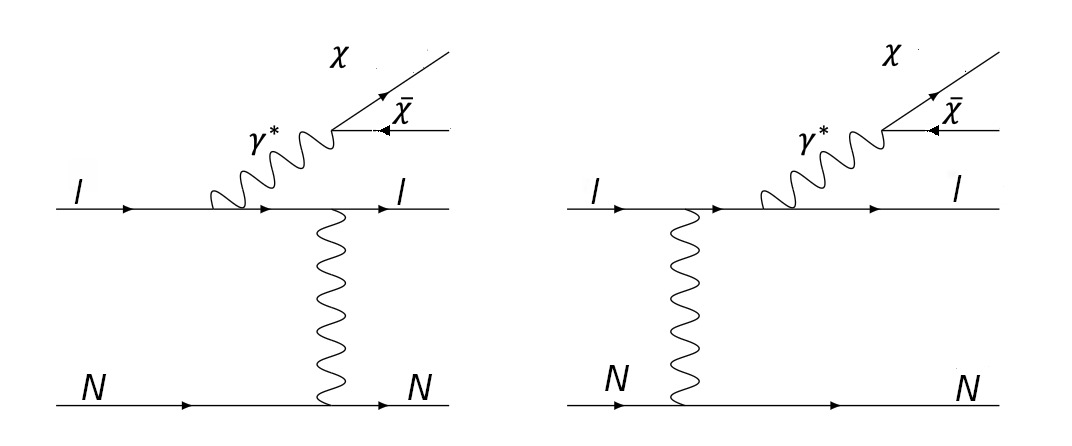}
\caption{Feynman diagrams describing  bremsstrahlung-like   signature for the MCPs pair production $l N \to l N \gamma^* (\to \chi \bar{\chi})$. 
\label{lNTolNChiChiDiagram} }
\end{figure}
 
 For probing millicharged particles  one can utilize a high-energy charged leptons, $l^-$, scattering of 
 nuclear targets, $N$, see Fig.~\ref{lNTolNChiChiDiagram}. In this case,  the MCPs search concept relies on the measuring of the  missing-energy signatures of the outgoing lepton
 from the reaction chain, involving off-shell SM photon $\gamma^*$ that emits hidden fermions 
  \begin{equation}
 l N\to l N \gamma^* \to l  N \chi \bar{\chi}. 
 \label{lNtolNChiChi22}
 \end{equation}
Part of the initial lepton energy can be carried away by the undetected pair of the millicharged particles.  

The  sensitivity of lepton missing momentum experiments
benefits from the quadratic dependence of the 
MCP signal yield on the mixing parameter ($N_{\rm sign} \propto \epsilon^2$), enabling exploration of the 
parameter space at $ \epsilon \lesssim 10^{-5} - 10^{-3}$ in the typical mass 
range $10~\mbox{MeV} \lesssim m_\chi \lesssim 1~\mbox{GeV}$
with relatively modest leptons-on-target (LOT) 
statistics of $10^{14} - 10^{16}$. This represents a significant advantage of the lepton missing momentum experiments over proton-based 
facilities with $N_{\rm sign}\propto \epsilon^4$ requiring substantially higher beam exposures at the level of 
$\mbox{POT} \simeq 10^{20}-10^{22}$, where POT is the number of protons 
accumulated on target~\cite{LSND:2001akn,Magill:2018tbb,MiniBooNE:2018esg,MicroBooNE:2017kvv}.
However, achieving the benefit of lepton missing momentum experiments necessitates 
suppressing the background rate to a sufficiently small value, on the order of $\lesssim \mathcal{O}(10^{-16})-\mathcal{O}(10^{-14})$ per incident lepton, where the irreducible SM background may play an important role~\cite{LDMX:2025bog}.

\begin{figure*}[!tbh]
\centering
\includegraphics[width=0.48\textwidth]{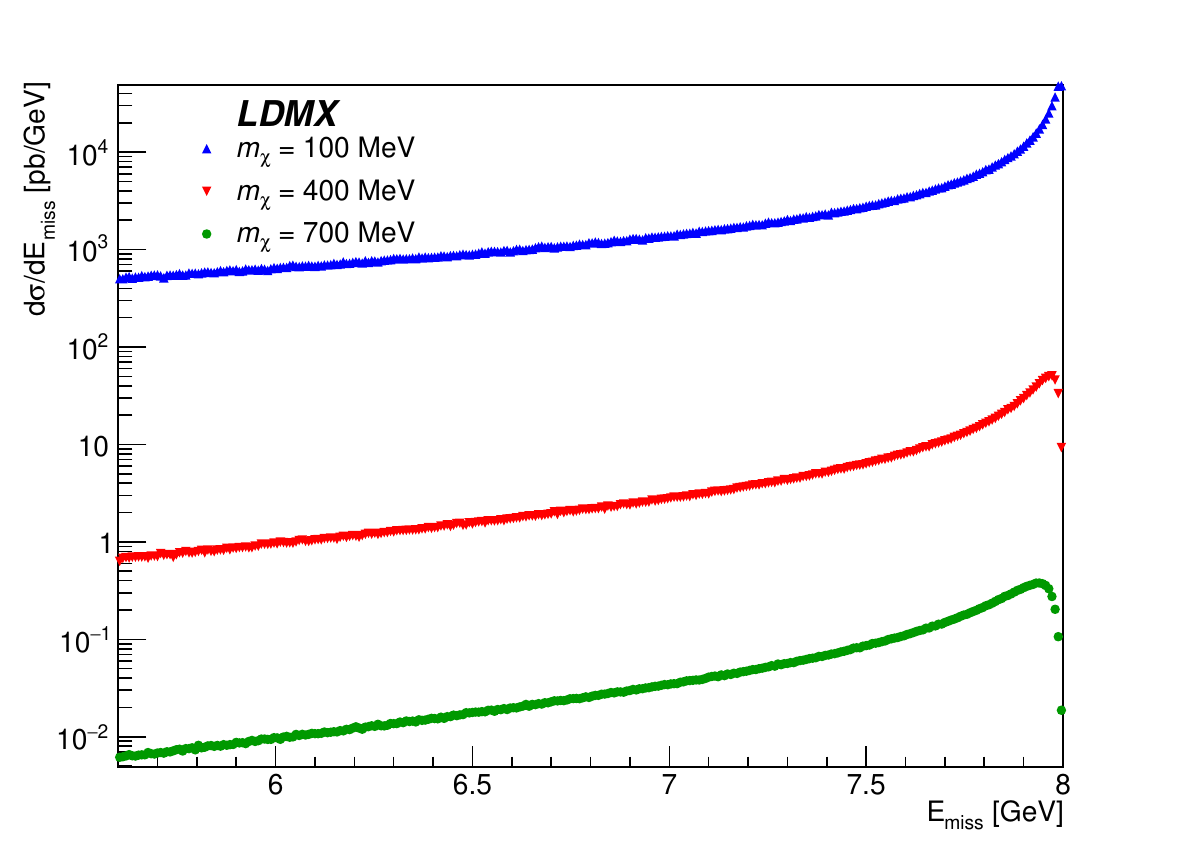}
\includegraphics[width=0.48\textwidth]{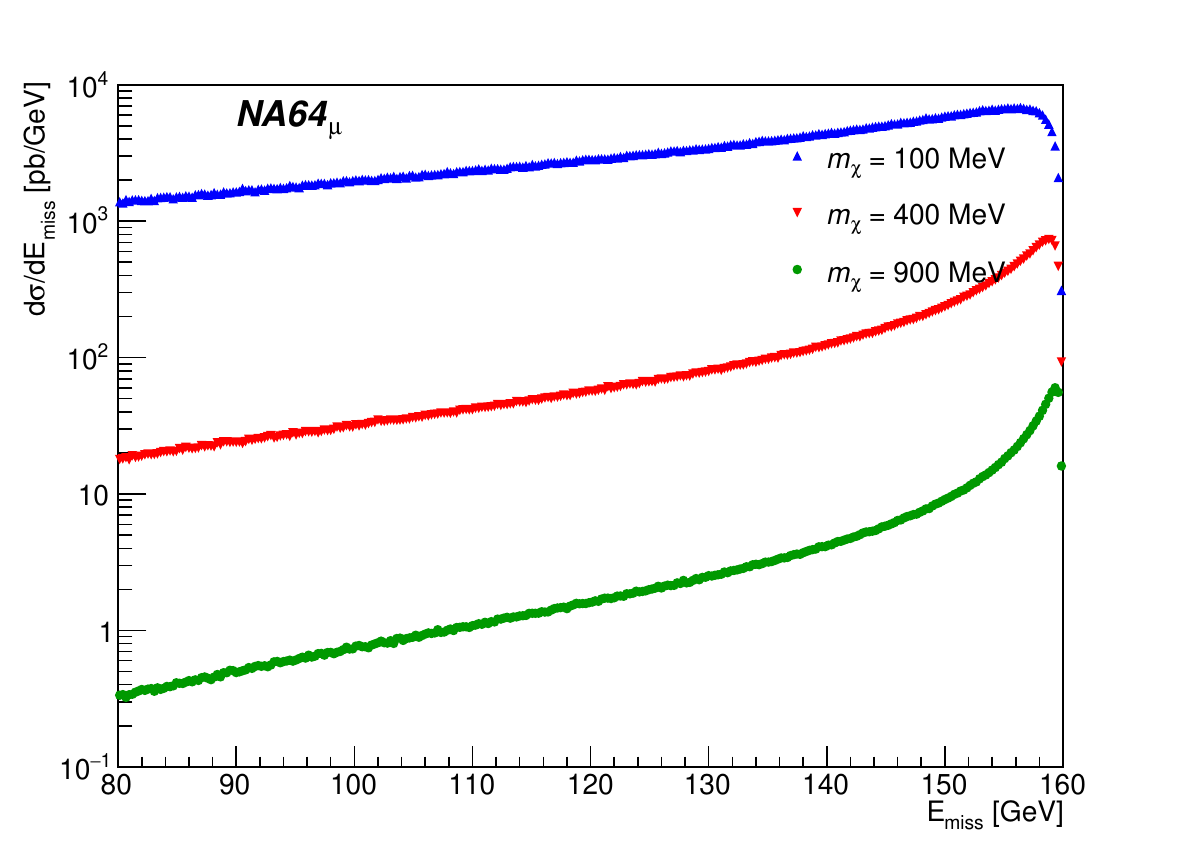}
\caption{ The differential cross sections of the process $l N \to l 
N  \gamma^* (\to \chi \bar{\chi})$
as function of the  missing energy $E_{\rm miss}=E_{\chi}+E_{\bar{\chi}}$  
 for the set of MCP masses.  
 Left panel shows the typical spectra for  LDMX experiment with electron beam energy of $E_{\rm e} \simeq 8~\mbox{GeV}$.  Right panel is the differential cross section for the NA64$\mu$ experiment, with muon beam energy of $E_\mu \simeq 160~\mbox{GeV}$.  We set $\epsilon=1$. 
\label{diffCSmcp}}
\end{figure*}

In order to calculate the bremsstrahlung-like  
production  cross section of  millicharged particles in the  laboratory, $l  N \to l  N \chi \bar{\chi}$, 
one can exploit either the exact tree-level (ETL) method~\cite{Arefyeva:2022eba} or modified Weizsaker-
Williams (WW) approach~\cite{Gninenko:2018ter}. The ETL method for $l N \to l N \chi \bar{\chi}$  
cross section calculation~\cite{Arefyeva:2022eba} is  the  straightforward phase space 
integration of the amplitude squared for the process $l N \to l N \chi \bar{\chi}$  without any 
approximation. Otherwise, exploiting the modified
WW approximation~\cite{Gninenko:2018ter} reduces  the number of phase space integrals  for the calcuation of the $l N \to l N \chi \bar{\chi}$ cross section and  implies the energy  of the incoming particle is  much greater 
than the particle masses, $E_l \gg m_l, m_\chi$ (for details see  e.~g.~Ref.~\cite{Kim:1973he} 
and  references therein).  So that the WW cross  section has to be considered as a simplified approximation 
of the ETL one, since the WW approach implies the factorization of the $ l N \to l N \chi \bar{\chi}$ cross 
section phase space into two terms, associated with off-shell Compton-like scattering,  
$l \gamma^*\to l \gamma^*$,  and off-shell photon emission of MCPs, $\gamma^*\to \chi \bar{\chi}$ (for 
details see   e.~g.~Ref.~\cite{Gninenko:2018ter} and  references therein). However, the modified 
WW method may result in a sizable cross section deviation  from 
the ETL one in the soft collinear  region, $E_\chi + E_{\bar{\chi}} \lesssim E_l$, where $E_\chi$ and 
$E_{\bar{\chi}}$ are the energies of the outgoing millicharged particles~\cite{Voronchikhin:2024vfu}.  
In what follows, the ETL method can be considered as a more reliable one.
The aim of the present paper is twofold.

First, we calculate the MCPs sensitivity of the muon missing momentum experiments NA64$\mu$
implying that the bremsstrahlung-like production cross section of the MCPs, 
$\mu N \to \mu N \gamma^* (\to \chi \bar{\chi})$
is calculated by exploiting the ETL method with the state-of-the-art CalcHEP routine~\cite{Belyaev:2012qa}.  The typical number of 
 muons accumulated on target for NA64$\mu$ is expected to be 
$\mbox{MOT}\lesssim 10^{14}$ by the final phase of running~\cite{NA64:2024nwj}.

Second, we implement the developed  ideas of MCP production cross in the framework of ETL method to estimate of the expected reach of the Light Dark Matter eXperiment (LDMX) 
for the millicharged particles in the  mass range of interest,  
$10~\mbox{MeV} \lesssim m_{\chi} \lesssim 1~\mbox{GeV}$.  Our analysis demonstrates promising 
sensitivity for MCP detection through two primary production mechanisms: 
(i) the reaction $e N \to e N \gamma^* (\to \chi \bar{\chi})$ yields MCP pairs via an 
off-shell photon, analogous to bremsstrahlung emission;
(ii)  MCPs may also originate from the invisible decays of short-lived vector mesons 
 produced in photo-nuclear interactions within the target. Quantitative calculations confirm the 
experimental feasibility of probing MCPs through these channels, highlighting LDMX’s 
potential to probe previously inaccessible regions of parameter space. The typical number of 
the electron accumulated on target for LDMX is expected to be 
$\mbox{EOT}\lesssim  \mathcal{O}(10^{16})$~\cite{Berlin:2018bsc,Schuster:2021mlr,LDMX:2025bog}.

The paper is organized as follows. In Sec.~\ref{ExperimentalBenchmark}, we discuss missing energy signal associated with MCPs 
production. In Sec.~\ref{NA64Cross-section}, we derive bremsstrahlung-like cross section of MCPs production. In Sec.~\ref{BremLimits}, we 
derive the MCPs limits for NA64$\mu$ and LDMX that arise from $l N \to l  N \gamma^* (\to \chi \bar{\chi})$ and $\gamma N \to N V (\to \chi \bar{\chi})$. 
 We conclude in~Sec.~\ref{Conclusion}. We also provide some helpful formulas in Appendix~\ref{BayesianUpperLimit}.

\section{The  missing energy signature
\label{ExperimentalBenchmark}} 

This section highlights the experimental characteristics of fixed-target facilities, such as 
LDMX~\cite{Mans:2017vej,Berlin:2018bsc,LDMX:2018cma,Ankowski:2019mfd,Schuster:2021mlr,Akesson:2022vza} or
NA64$\mu$~\cite{Gninenko:2014pea,Gninenko:2018tlp,Kirpichnikov:2021jev,Sieber:2021fue,NA64:2024nwj,NA64:2024klw} 
which are designed to probe invisible signatures via lepton missing-energy processes of the form:
\begin{equation}
l N \to l N + E_{\rm miss}\,,
    \label{generalMissinEnergyProcess}
\end{equation}
where label $l=(e,\mu)$ denotes the incident lepton beam (electrons for LDMX and muons for NA64$\mu$). As 
previously discussed, the missing energy $E_{\rm miss}$  arises from the production 
of  $\chi \bar{\chi}$ through the off-shell photon  $\gamma^*$, such that, 
$E_{\rm miss } = E_{\chi }+ E_{\bar{\chi}}$. 

To quantify the experimental sensitivity, we estimate the expected number of missing-energy events for a lepton 
beam incident on a fixed target:
\begin{equation}
N_{\gamma^*\to \chi \bar{\chi}} \simeq \mbox{LOT} \times \frac{\rho N_A}{A} L_T \times \int\limits^{x_{\rm max}}_{x_{\rm th}}  dx  \frac{d \sigma_{2\to 4} }{dx}(E_l),
\label{NumberOfMissingEv1}
\end{equation}
where $E_l$ is the initial energy of the beam, $A$ and $\rho $ are the atomic weight and density of the target material, respectively,
$N_A$ is Avogadro's number, $\mbox{LOT}$ is the total number of 
leptons  on target,  $L_T$ is the effective target thickness, $d \sigma/dx$ is the differential cross section for the process (\ref{generalMissinEnergyProcess}), and 
$x_{\rm th}$ defines the kinematic range of the missing-energy fraction  $x=E_{\rm miss}/E_l$ determined  by the experimental setup. We note that typically $x_{\rm max} \simeq 1$ in Eq.~(\ref{NumberOfMissingEv1}) for both LDMX and NA64$\mu$.  

One of the most extensively investigated production mechanisms involves dark bremsstrahlung mediated by a vector 
boson (SM photon or hidden massive vector). However, the authors of Ref.~\cite{Schuster:2021mlr} examine an alternative source of electron missing  energy-momentum signatures, wherein bremsstrahlung photons $\gamma$ undergo conversion into high-energy vector mesons $V=( \rho, \omega, \phi )$, via 
exclusive photoproduction processes, $\gamma^* N \to N V  (\to \chi \bar{\chi})$. These mesons subsequently decay into millicharged particles, dark matter particles or other non-detectable final states, including neutrinos.

The typical number of the vector meson, $V$ produced via exclusive photoproduction can be expressed as follows
\begin{equation}
N_V = \mbox{EOT} \times f_{\rm brem} \times P_V,
\end{equation}
where  $f_{\rm brem}$ is a fraction of bremsstrahlung photons~\cite{Schuster:2021mlr} produced by electrons,
and $P_V$ is typical probability that the photon  undergoes an exclusive photoproduction process.
The latter can be expressed as follows:
\begin{equation} \label{PVexpression}
P_V \simeq \frac{9}{7} \frac{\sigma_0^V X_0 f_{\rm nuc}^V}{m_p}, 
\end{equation}
where $X_0$ is the radiation length, $\sigma_0^V$ is the cross section for exclusive
photoproduction on a single nucleon, and $f_{\text{nuc}}^V$ is an order-one correction factor.
  Authors of Ref.~\cite{Schuster:2021mlr} estimated the typical factor at the level of $f_{\text{brem}} = 0.03$,
 for LDMX Phase II design parameters,  that implies the meson yield  to be in the range from 
 $\mathcal{O}(10^9)$ to $\mathcal{O}(10^{10})$. In particular, for the typical ultimate number of electrons $\mbox{EOT}\simeq 2 \times 10^{16}$, one can estimate the vector meson yield~\cite{Schuster:2021mlr}
\begin{equation}
    N_\rho \simeq 6\times 10^{10}, \quad N_\omega \simeq 6\times 10^{9}, \quad N_\phi \simeq 10^{9}.
    \label{NimberOfMesonsLDMXultimate}
\end{equation}
As a result, the  signal yield associated with invisible decay into pair of millicharged particles reads
\begin{equation}
    N_{V\to \chi \bar{\chi}} \simeq N_{V} \times \mbox{Br} (V\to \chi \bar{\chi}),
    \label{NumbSignEVMesDec}
\end{equation}
where $ \mbox{Br} (V\to \chi \bar{\chi})$ is the invisible branching fraction  of vector meson
associated with MCP scenario. Specifically, it is given by
\begin{align}
 & \mbox{Br}(V\to \chi \bar{\chi}) \simeq  \epsilon^2 \times \mbox{Br}(V\to e^+e^-)  \times  \label{BrVtoChiChi}\\
 & \times \left(1+2m_{\chi}^2/m_{V}^2\right) \left(1-4 m_{\chi}^2/m_V^2\right)^{1/2}\,,   \nn 
\end{align}
where $m_V$ is a typical mass of the vector meson and $\mbox{Br}(V\to e^+e^-)$
is a branching fraction of visible meson decay into to $e^+e^-$ pair. 
In Eq.~(\ref{BrVtoChiChi}) we imply that $m_\chi \gg  m_e$ and $m_V \gg m_e$.  

\begin{figure}[!tbh]
\centering
\includegraphics[width=0.5\textwidth]{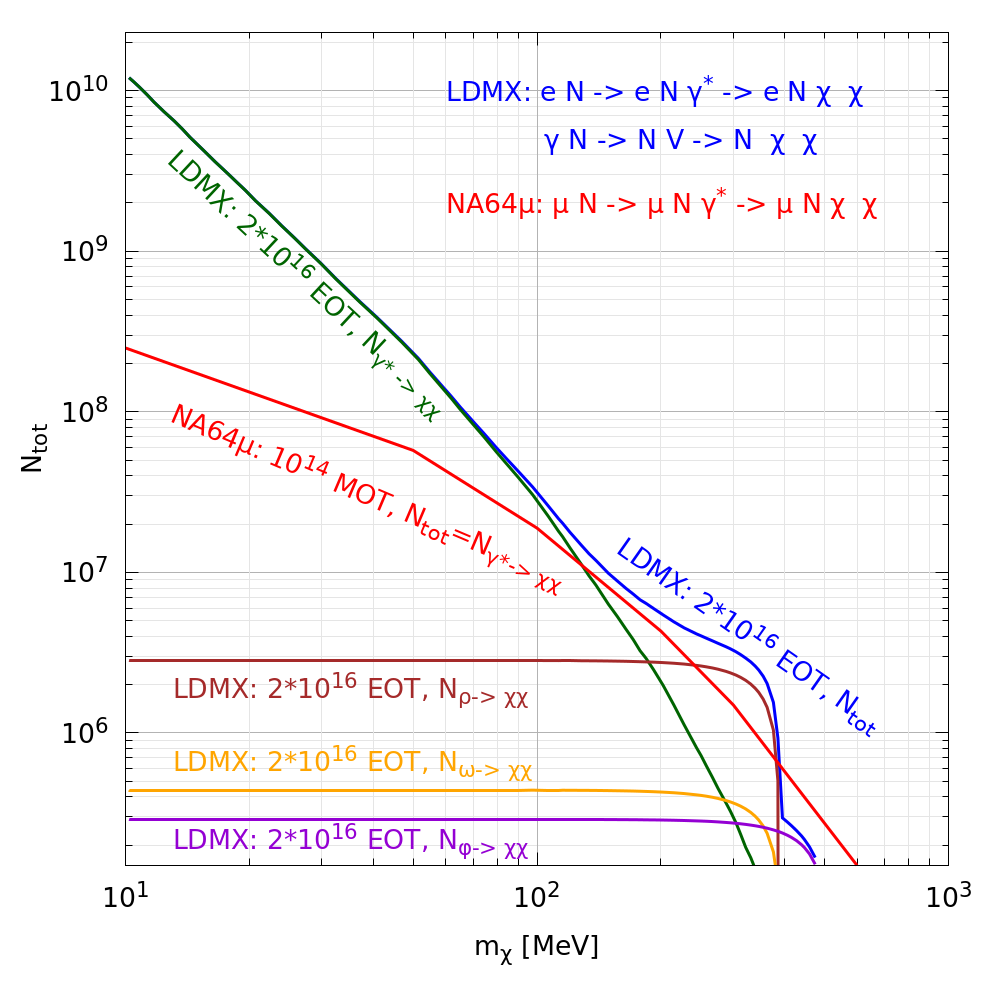}
\caption{The plot shows the total number of mCPs produced versus their mass for both NA64$\mu$ and LDMX, with the charge fraction chosen to be $\epsilon=1$. The MCPs are produced in bremsstrahlung-like reactions and various meson decays. At the NA64$\mu$ facility, the dominant production mechanism is  virtual photon emission $\gamma^* \to \chi \bar{\chi}$ (red solid line). 
At the LDMX experiments, the MCPs are produced through the bremsstrahlung-like channel $\gamma^*\to \chi \bar{\chi}$ (dark green), rho meson  
$\rho \to \chi \bar{\chi}$ (brown solid line), omega meson $\omega \to \chi \bar{\chi}$ (yellow solid line), and phi meson decay $\phi \to \chi \bar{\chi}$ (violet solid line). The resulted number of MCPs produced at the LDMX is also shown (blue solid line). 
\label{NtotFig} }
\end{figure}

We emphasize that  for the NA64$\mu$ experiment, the  number of produced mesons is 
negligible and does not contribute  significantly to the missing energy signal yield 
due to fewer number of  bremsstrahlung like photons produced by incident muons, 
$f^{{\rm NA64}\mu}_{\rm brem.}  \ll f^{\rm LDMX}_{\rm brem.}$. Specifically, the numerical estimates yield $f^{{\rm NA64}\mu}_{\rm brem.} \simeq 7\times 10^{-4}$, 
implying that NA64$\mu$ total number of signal events is
\begin{equation}
N_{\rm tot}\simeq N_{\gamma^*\to \chi \bar{\chi}}.
\label{TotSignNA64mu}
\end{equation}
However, for the LDMX facility, the resulted number of missing energy events reads 
\begin{equation}
N_{\rm tot} \simeq N_{\gamma^*\to \chi \bar{\chi}} + \sum_{V = \rho, \omega, \phi} N_{ V \to \chi \bar{\chi} }. 
\label{TotSignLDMX}
\end{equation}
In Fig.~\ref{NtotFig} we show the total number of millicharged particles produced 
at NA64$\mu$ and LDMX as a function of~$m_{\chi}$. The kink in signal observed for 
the LDMX case at $m_\chi \simeq 400~\mbox{MeV}$ is due to the contribution from the 
vector meson decays to the total yield.

The sensitivity estimates for both NA64$\mu$ and LDMX used in our paper are based on the detailed 
studies reported, respectively    Ref.~\cite{NA64:2018iqr} and Ref. \cite{LDMX:2025bog}. The 
experimental signatures of  MCP and $A'$ (dark photon) production are quite similar, both LDMX 
and NA64mu adopt the similar search strategy by looking for  events with a single recoil 
primary beam particle in the final state accompanied by a large missing energy, typically  
$\gtrsim 50\%$ of the primary beam energy. The later is a specific criteria used for the 
powerful suppression of background events. To fulfill this criteria the construction of a  
fully hermetic detector is required. 

\subsection{The LDMX experiment} 

LDMX is a planned fixed-target experiment,  utilizing an electron beam to search for dark sector 
particles. A key feature of LDMX is its precise measurement of electron missing momentum, enabling it to probe 
the process described in Eq.~(\ref{lNtolNChiChi22}). This approach provides complementary sensitivity compared to 
other electron fixed-target experiments such as 
NA64e~\cite{Gninenko:2017yus,Gninenko:2019qiv,Banerjee:2019pds,NA64:2021xzo,Andreev:2021fzd,NA64:2023wbi,NA64:2022yly}. 
The LDMX  employs stringent missing energy selection criteria and sophisticated veto 
systems,  resulting in experiment with negligible background 
contamination~\cite{Berlin:2018bsc}.

The LDMX detector configuration comprises a thin target, a high-resolution silicon tracker, and both 
electromagnetic and hadronic calorimeters (see, e.g., Refs.~\cite{Berlin:2018bsc,Akesson:2022vza} for 
more detailed descriptions). The experimental signature relies on the production of dark sector 
particles in the upstream target, leading to significant energy loss from the primary electron 
beam. The missing momentum is reconstructed using the silicon tracker in conjunction with energy 
measurements from the downstream calorimeters. The analysis requires a reconstructed electron 
energy $E_{\rm e}^{\rm rec} \lesssim 0.3 E_{\rm e}$, corresponding to a threshold value of $x_{\rm th} =0.7$  in Eq.~(\ref{NumberOfMissingEv1}). 

For this study, we consider an aluminum target (Al) with the following properties: 
radiation length $X_0=8.9~\mbox{cm}$, density $\rho=2.7\,\mbox{g cm}^{-3}$, atomic mass 
$ A=27\,\mbox{g mole}^{-1}$, and atomic number $ Z=13$. The target thickness is set to 
be $L_T \simeq 0.4 X_0$. LDMX anticipates accumulating approximately 
$\mbox{EOT}\simeq \mathcal{O}(10^{16})$ electrons-on-target  with a beam energy up to 
$E_{\rm e}= 8\,\mbox{GeV}$~(see 
e.g.~Ref.~\cite{Akesson:2022vza,Berlin:2018bsc}  and references therein).  

The projected background for the LDMX experiment originates 
from several key sources~\cite{LDMX:2025bog}: (i) unbiased electron interactions, (ii) photo-nuclear interactions, and 
(iii) electro-nuclear  interactions, (iv) muon conversions. To maintain sensitivity at a proposed 
exposure of $\mbox{EOT} \lesssim 2\times 10^{16}$, a program 
of  front-end electronics development is 
planned~\cite{LDMX:2025bog}. We expect that the implementation of these 
upgrades will be designed to suppress the reducible background to a 
level  of $b \lesssim 1$ event.

\begin{figure}[!tbh]
\centering
\includegraphics[width=0.5\textwidth]{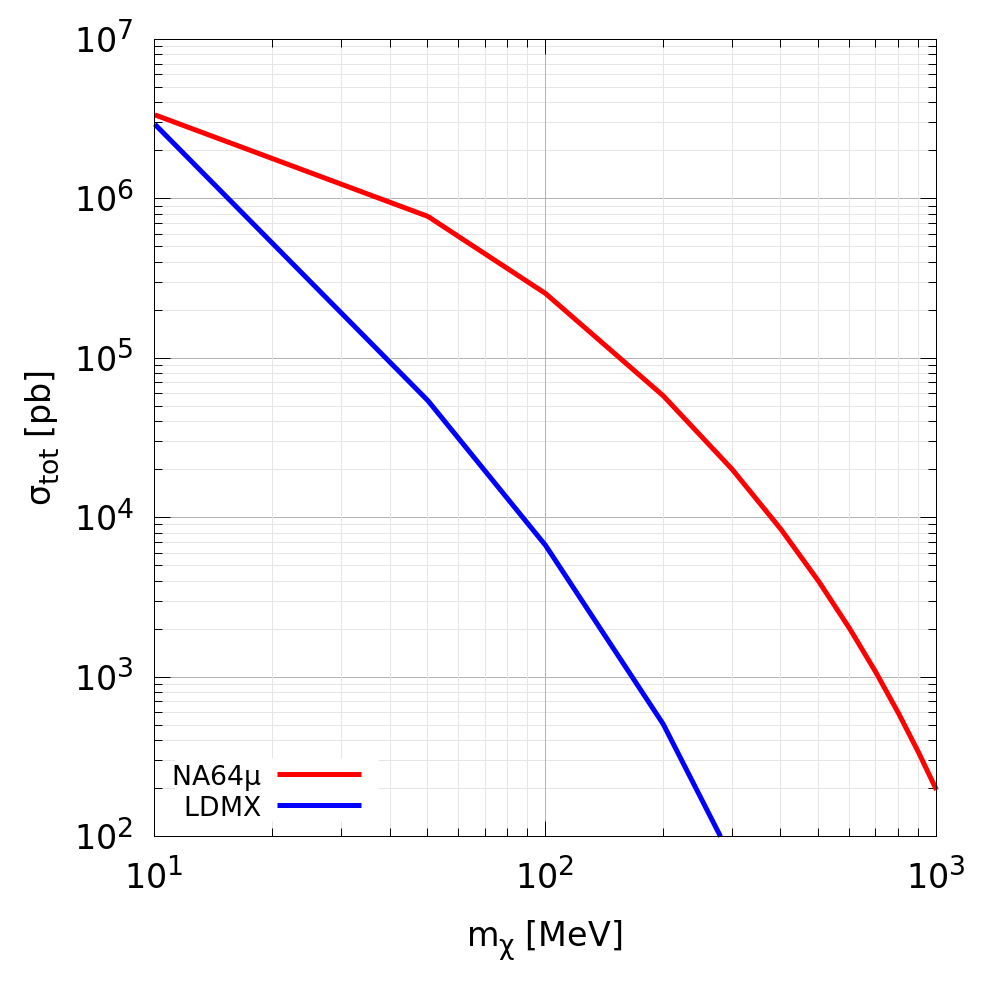}
\caption{Total cross section of MCPs production as a function of $m_{\chi}$ for both the lepton missing momentum experiments, implying that $\epsilon=1$. 
We integrate the cross sections  
over the experimental cut range  $x_{\rm th} \lesssim x \lesssim x_{\rm max}$. Red solid line is the 
cross section for the NA64$\mu$
 experiment, blue solid line corresponds to the LDMX fixed target facility, 
\label{NA64TotCS} }
\end{figure}

\subsection{ The NA64$\mu$ experiment}
The NA64$\mu$ experiment~\cite{Kirpichnikov:2021jev,Sieber:2021fue} is a fixed-target facility designed 
to investigate dark sector particles through muon-nucleon interactions, particularly via the process   
$\mu N \to \mu N \gamma^* (\to \chi \bar{\chi})$. For our analysis, we assume a muon beam energy of 
$E_\mu \simeq 160~\mbox{GeV}$ and an integrated muon flux of approximately $\mbox{MOT}\lesssim 10^{14}$, 
corresponding to the projected experimental statistics  by the final phase of running.

The detector configuration employs a lead-based shashlyk electromagnetic calorimeter, which simultaneously serves 
as the target material. The target thickness is set to $L_{T}\simeq 40 X_0\simeq 22.5\, \mbox{cm}$, where 
$X_0\simeq 0.5~\mbox{cm}$ represents the radiation length of the lead target (Pb) ($\rho=11.35\,\mbox{g cm}^{-3}, A=207\,\mbox{g mole}^{-1}, Z=82$).
Following previous studies~\cite{Chen:2017awl,Gninenko:2018ter}, we neglect energy loss due to muon 
stopping in the  lead target, as the average energy attenuation $\langle dE_\mu/dx \rangle \simeq 12.7 \times 10^{-3} \mbox{GeV}/\mbox{cm}$ is negligible 
 for 
ultra-relativistic muons with the energy $E_\mu \simeq 160\, \mbox{GeV}$. 

NA64$\mu$  incorporates two magnetic spectrometers to enable precise momentum reconstruction for both 
incoming and outgoing muons~\cite{Sieber:2021fue}. Our analysis implements a kinematic cut on the recoiling muon 
energy of $E_\mu^{\rm rec} \lesssim 0.5E_\mu \simeq 80\,\mbox{GeV}$, corresponding to a threshold parameter $x_{\rm th} =0.5$ in the missing energy analysis. In the present conservative analysis we neglect
muon recoil angle, implying that it is sufficiently small to impact the sensitivity of NA64$\mu$.     

For NA64$\mu$, the main background sources 
are~\cite{NA64:2024nwj}: (i) muon momentum mis-reconstruction; (ii) hadron in-flight decays; (iii) single-hadron punch-through; (iv) dimuons 
production; (v) detector non-hermeticity. 
We expect that  rejection of this background for NA64$\mu$ with 
$\mbox{MOT} \lesssim 10^{14}$ is foreseen through 
electronic upgrades, 
which are expected to suppress the number of such events to~$b \simeq 1$.

\section{The bremsstrahlung-like cross section
\label{NA64Cross-section}}

This section outlines the methodology for calculating the cross section of 
millicharged particle  production in lepton-nucleus scattering. 
The corresponding Feynman diagrams for the tree-level amplitude are illustrated in 
Fig.~\ref{lNTolNChiChiDiagram}. It is worth noticing, that  the dominant contribution arises from amplitude terms linear in the 
millicharge parameter $\epsilon$, while higher-order trident contributions proportional to $\epsilon^2$ are neglected (see e.~g.~Ref.~\cite{Gninenko:2018ter} for detail).

The squared tree-level amplitude is integrated over the phase space of the process in Fig.~\ref{lNTolNChiChiDiagram} using the latest version of the CalcHEP software package~\cite{Belyaev:2012qa}. The nucleus-photon interaction vertex contains elastic form factor $F(-q^2)$ 
\begin{equation}
\label{FF1}
i e Z F(-q^2) \overline{U}_N \gamma_\mu U_N, 
\end{equation}
depending on the time-like momentum transfer squared $q^2 < 0$ as
\begin{equation}
\label{FF2}
F(-q^2) =  \frac{ (-a^2 q^2)}{(1 - a^2 q^2)} \frac{1}{(1 - q^2/d)} \,.
\end{equation}
The screening parameter $a$ and the nucleus scale parameter $d$ are given by $a=111 Z^{-1/3}/m_e$ and  $d=0.164 A^{-2/3}\, \mbox{GeV}^2$,  respectively

For each target nucleus (defined by its chemical composition) and a given lepton beam energy $E_l$, we 
compute the total production cross section $\sigma_{\rm tot}$ as a function of the MCP mass $m_\chi$, 
scanning the range $10\, \mbox{MeV} \lesssim  m_{\chi }\lesssim 1\, \mbox{GeV}$.
The phase-space integration is implemented via the  VEGAS Monte Carlo algorithm, 
employing $N_{\rm session} =10$ independent optimization runs with $N_{\rm calls}=10^6$ evaluations per 
run. The adaptive grid refinement in VEGAS achieves a numerical convergence at the level of $0.1\%- 0.01\%$ precision throughout the  CalcHEP computation.

The differential cross sections as a function of the missing energy are presented in Fig.~\ref{diffCSmcp} for the signal box  region $E_l^{\rm th} \lesssim E_{\rm miss} \lesssim E_l$, where $E_l^{\rm th}= x_{\rm min} E_l$. 
These results are obtained from fixed-target experiments for different dark matter masses 
$m_{\chi}$. Both the NA64$\mu$ and LDMX experiments exhibit a  peak in the 
differential cross sections near $E_{\rm miss}\simeq E_l$. This observation suggests that 
the signal is strongly forward-peaked when $E_{\rm miss} \gg m_{\chi}$, indicating that the 
majority of the beam energy is transferred to the millicharged pair.

In the case of the muon beam cross sections measured by NA64$\mu$, the peak observed at missing energies $E_{\rm miss} \simeq E_\mu$ for $m_\chi \simeq 100~\mbox{MeV}$ is 
suppressed. This mitigation occurs because the production rates of millicharged particles imply a soft bremsstrahlung-like regime, 
particularly when the MCP mass is small compared to the muon mass, $m_\chi \lesssim m_\mu$.  

In Fig.~\ref{NA64TotCS}, the total cross sections for the NA64$\mu$ and LDMX experiments are 
presented. Notably, the NA64$\mu$ total cross section (for a lead (Pb) target, $Z=82$, with a 
muon beam energy $E_\mu=160\, \mbox{GeV}$) is comparable in magnitude to the LDMX total cross
section (for an electron beam energy $E_{\rm e}=8\, \mbox{GeV}$) for small masses of the millicharged 
particles $m_\chi \simeq 10~\mbox{MeV}$.

This implies a boost in the cross section for electron beams over muon beams in the 
low-mass region 
$m_\chi \lesssim 10~\mbox{MeV}$, provided the electron beam flux is sufficiently 
high. However, $\sigma_{\rm tot}^e$ falls 
off more steeply than $\sigma_{\rm tot}^\mu$ as  $m_\chi $ approaches 
$1~\mbox{GeV}$. Consequently, for experiments with comparable lepton 
statistics on target, a muon beam of larger energy becomes more favorable
in the higher-mass region  $m_\chi \lesssim 1~\mbox{GeV}$.

We emphasize that for the parameter space of interest, specifically for the masses 
\(m_\chi \gtrsim 10~\text{MeV}\) and sufficiently small energies of the millicharged 
particles,  \(E_{\text{miss}} \equiv E_{\chi} + E_{\bar{\chi}} \ll E_l\), the cross 
section calculation procedure using CalcHEP is free from collinear and infra red singularity issues. This 
absence of singularities is justified by the effective convergence of the VEGAS 
integration algorithm, which is achieved at a precision level of \(0.1\% - 0.01\%\). One 
might anticipate that the peak forward region, characterized by a large total energy of 
the millicharged particles, \(E_{\text{miss}} \simeq E_l\), would require special 
treatment in the calculation of the differential cross section. However, this is not the 
case due to the excellent performance of the CalcHEP integration algorithm.

For sufficiently small masses, specifically millicharged particles  with 
\( m_{\chi} \ll m_e \), the VEGAS algorithm in CalcHEP requires special treatment. This 
is because very light MCPs (e.~g.,~\( m_{\chi} \lesssim 1~\text{MeV} \)) have a production 
cross section that is dominated by the infra-red regime, where most particles are 
produced with relatively small energy, \( E_{\text{miss}} \ll E_l \). In this limit, the 
differential cross section of the MCP production peaks at \( E_{\text{miss}} \ll E_l \), causing the VEGAS 
algorithm to fail to converge with the desired accuracy.  The  latter implies 
double-precision floating-point format for the energies, masses and Mandelstam variables, such 
as transfer momentum squared $t = -q^2$. To resolve this issue, one must use 
quadruple-precision floating-point format for the cross section parameters in the VEGAS 
Monte-Carlo  integration~\footnote{Private communication with CalcHEP developer Alexander 
Pukhov}. However, the present paper focuses only on larger millicharged masses, 
\( m_{\chi} \gtrsim 10~\text{MeV} \), where these issues do not occur.

In the following analysis, we neglect the signal from photon emission on the nuclear leg. This neglection is justified by the  analysis in Ref.~\cite{Chu:2018qrm}, which shows that the relevant signal is parametrically smaller by a factor of
$$  
\frac{\sigma_{N\to N \gamma^*}}{\sigma_{l \to l \gamma^*}} \simeq 
\left( \frac{\alpha^4 Z^4 \epsilon^2 }{m_{N}^2}
\right)\times
\left(  \frac{\alpha^4 Z^2 \epsilon^2 }{m_{l}^2}
\right)^{-1}  \simeq \left( \frac{m_l Z }{m_N} \right)^2. 
$$
As a result, this yields the suppression factors on the order of $6\times10^{-8}$ and $10^{-3}$ for 
LDMX and NA64$\mu$, respectively.  Moreover, such corrections provide a negligible contribution to the expected reach of the lepton missing momentum  facilities of interest.

The MCP signal selection cuts on the MCP signal were studied in Ref.\cite{NA64:2018iqr}, see 
also \cite{Chu:2018qrm,Gninenko:2014pea,Sieber:2021fue,NA64:2024klw} by maximizing the signal sensitivity versus the expected level of 
background. The optimal sensitivity was found slightly dependent on the cut 
$E^{\rm rec}_\mu\simeq 80~\mbox{GeV}$  
 variation within $\pm~5~\mbox{GeV}$. The typical precision of the muon momentum 
 reconstruction including multiple scattering in the target,  is at the level of $\simeq3\%$ 
 for the momentum $p_\mu \simeq 80~\mbox{GeV}$, and does not affect the sensitivity of the 
 search much. The dominant effect is expected  from the primary muon momentum 
 miss-measurements, e.~g.~due to secondary delta-electron emission along the track.   
 
    For the LDMX experiment we relay mostly on the cuts used in the studies performed in 
    Ref.~\cite{Schuster:2021mlr,Akesson:2022vza,Berlin:2018bsc} for the search for invisible 
    decay of the dark photon $A'$ in the sub-GeV mass region, as the experimental signatures 
    of the MCP and the $A'$ are the same.  This is also justified by the fact that their  
    differential energy spectra  have similar shape due to kinematics of the MCP and $A'$ 
    production with the dominant yield from the forward peak region.

\begin{figure}[!tbh]
\centering
\includegraphics[width=0.5\textwidth]{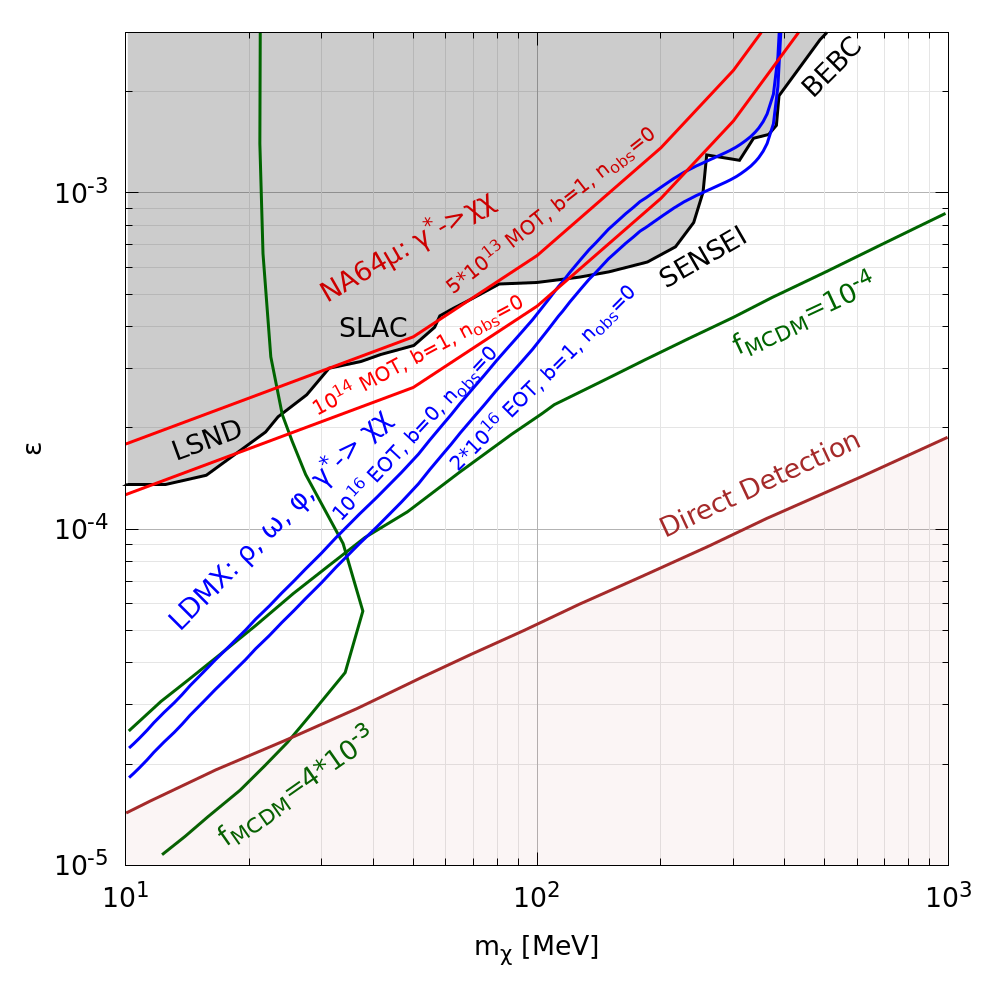}
\caption{The expected experimental reach of the NA64$\mu$ (red solid lines) and LDMX (blue 
solid lines) fixed target facilities in the $(\epsilon, m_\chi)$ plane. For the LDMX 
experiment, we take into account invisible decays of vector mesons to MCPs, $V\to \chi \bar{\chi}$, and MCP production by energetic beam electrons via the bremsstrahlung-like 
channel $\gamma^*\to \chi \bar{\chi}$. The projected limits, for an ultimate prospect 
statistics of $\mbox{EOT}=2\times 10^{16}$ and $10^{16}$ in the MCP mass range
$10~\mbox{MeV}\lesssim m_{\chi} \lesssim 1~\mbox{GeV}$, correspond to backgrounds of $b\simeq 1$ 
and $b=0$, respectively. We also show the expected reach of NA64$\mu$ for  prospect statistics of $\mbox{MOT}= 10^{14}$ and $5\times 10^{13}$, implying the virtual photon dominated channel of MCP production, $\gamma^* \to \chi \bar{\chi}$.
Both of these scenarios  imply a finite background of $b\simeq 1$ for the NA64$\mu$.
 The limits based on data of LSND~\cite{LSND:2001akn}, SLAC~\cite{Prinz:1998ua}, BEBC~\cite{Marocco:2020dqu}, and SENSEI~\cite{SENSEI:2023gie} are shown by grey shaded region (these bounds don't imply the MCP to be a DM or its fraction).  All constraints are set at 95\%\,CL.
 The shaded brown region is excluded by present direct detection  experiments as shown in Ref.~\cite{Emken:2019tni} (see text). 
The
green solid lines correspond to the millicharged dark matter explaining the EDGES anomaly Ref.~\cite{Liu:2019knx}. 
We emphasize, that our sensitivity of LDMX for bremsstrahlung like channel $e N \to e N \gamma^* (\to \chi \bar{\chi})$ is in a reasonable agreement with previous study of the authors Ref.~\cite{Berlin:2018bsc}. 
\label{MesonNA64Limits}}
\end{figure}

\section{The Expected reach
\label{BremLimits}}
In this section, we discuss the expected sensitivities of the NA64$\mu$ and LDMX fixed-target experiments to MCP production. This includes production via bremsstrahlung-like emission, $\gamma^* \to \chi \bar{\chi}$, and through vector meson decays, $V \to \chi \bar{\chi}$.
Using Eqs.~(\ref{TotSignLDMX}) and (\ref{TotSignNA64mu}) to estimate the number of produced MCP pairs, we derive the expected exclusion limits on the coupling constant $\epsilon$. We require $N_{\rm tot} \gtrsim s_{\rm up}$, which corresponds to the 95\% confidence level  upper limit on $\epsilon$ for a null experimental result, $n_{\rm obs}=0$, implying that zero signal events will be observed.

The detailed background evaluation presented in this work  relies on analysis of  
Refs.~\cite{NA64:2024klw,LDMX:2025bog}, where it is expected that background will be suppressed to the level $b \lesssim \mathcal{O}(1)$ by the final phases of experimental running.  
Based on this, in what follows we consider two benchmark cases for the conservative background estimates:
(i) a negligible background, $b\ll 1$, where it is sufficiently suppressed given the expected statistics, and
(ii) a non-zero background case, where it is constrained to $b=1$ for the anticipated statistics.
The details of the upper limit calculation  $s_{\rm up}$ can be found in Appendix~\ref{BayesianUpperLimit}.
It is also assumed that detector efficiencies are $100\%$  in both LDMX and NA64 muon.

In Fig.~\ref{MesonNA64Limits} we show the expected experimental reaches of NA64$\mu$ 
and LDMX for  $\mbox{MOT}\simeq 10^{14}$ and $\mbox{EOT}\simeq 2 \times 10^{16}$, respectively, implying non-negligible background, $b\simeq 1$.  In Fig.~\ref{MesonNA64Limits}  we  show  the NA64$\mu$ limits for  mitigated 
number of muons accumulated on target $\mbox{MOT}\simeq 5 \times 10^{13}$, implying the finite background 
$b\simeq 1$. In addition, for the suppressed background, $b=0$, we depict in Fig.~\ref{MesonNA64Limits} the 
expected reach of the LDMX for $\mbox{EOT}\simeq 10^{16}$. 

Note 
that  projected  limits on $\epsilon$ from LDMX fixed target facility 
are fairly strong $2\times 10^{-5} \lesssim  \epsilon \lesssim 5\times 10^{-4}$
for  the mass range of interest $10~\mbox{MeV} \lesssim m_\chi \lesssim 100~\mbox{MeV}$, even though 
the cross sectionof $\chi \bar{\chi}$ pair production at LDMX is smaller than the NA64$\mu$ one   
(see, e.~g.~Fig.~\ref{NA64TotCS} for detail).   
The enchantment of the  LDMX sensitivity   can be explained by 
the  projected accumulated statistics, 
$\mbox{EOT}\simeq 2\times10^{16}$, by the final phase  of  experimental  running. Nevertheless, 
the projected statistics of NA64$\mu$ at the level of $ \mbox{MOT} \simeq 10^{14}$ will allow to probe still 
unexplored region of the MCP parameter space $10^{-4} \lesssim  \epsilon \lesssim 5\times 10^{-4}$ in the mass 
range $10~\mbox{MeV}\lesssim  m_\chi \lesssim  100~\mbox{MeV}$. To note, 
the expected bounds of NA64$\mu$  
for $100~\mbox{MeV}\lesssim  m_\chi \lesssim  1~\mbox{GeV}$ have been already ruled 
out by  SENSEI~\cite{SENSEI:2023gie} and BEBC~\cite{Marocco:2020dqu} collaborations.

Let us discsuss an evaluation of vector meson contributions, specifically $V=( \rho, \omega, \phi)$, to millicharged particle  production in the LDMX experiment. Following the methodology established in \cite{Schuster:2021mlr}, we calculate the expected vector meson yield from photo-nuclear interactions $\gamma N \to N V$ for the anticipated total statistics of $\mbox{EOT}\simeq 2\times 10^{16}$. 

In the absence of signal events, it implies the approximate limit on $\epsilon$ for $\rho$ meson mode, 
$\rho\to \chi \bar{\chi}$ with $m_\chi \simeq m_\rho/2$, of the following form
\begin{align}
& \epsilon \gtrsim N_{\rho}^{-1/2} \cdot s_{\rm up}^{1/2} \cdot    \left(\mbox{Br}(\rho\to e^+e^-) \right)^{-1/2} 
\times 
\\
& \times  \left(1+2m_{\chi}^2/m_{\rho}^2\right)^{-1/2}
\left(1-4 m_{\chi}^2/m_\rho^2\right)^{-1/4}\,, \nn
\end{align}
Fig.~\ref{MesonNA64Limits} displays the resulting exclusion limits in the $(\epsilon, m_{\chi})$ parameter space. 
Notably, a viable discovery region emerges for  $m_\chi\lesssim 400~\mbox{MeV}$ and $\epsilon\simeq 10^{-3}$ which LDMX could probe through the $\rho \to \chi \bar{\chi}$ decay channel for~$\mbox{EOT}\simeq 2\times 10^{16}$.

To conclude this section, we discuss the direct detection (DD) constraints presented in Fig.~\ref{MesonNA64Limits}. The DD  bounds shown are derived from a combined analysis of data from SENSEI~\cite{Crisler:2018gci,SENSEI:2019ibb}, CDMS-HVeV~\cite{SuperCDMS:2018mne}, XENON10~\cite{Essig:2017kqs,XENON10:2011prx}, XENON100~\cite{Essig:2017kqs,XENON:2016jmt}, and DarkSide-50~\cite{DarkSide:2018ppu}, following the methodology of Ref.~\cite{Emken:2019tni,Anchordoqui:2021ghd}.

These bounds were computed under the assumption that MCPs constitute a subcomponent of DM with 
a fractional relic abundance of approximately \( f_{\rm MCDM} \simeq 4 \times 10^{-3} \).  The 
attenuation of a potential DM flux, due to scatterings on electrons and nuclei within the 
Earth's crust, atmosphere, and experimental shielding materials, significantly reduces the 
expected signal rate at a terrestrial detector for sufficiently large interaction cross-
sections. Consequently, DD experiments lose sensitivity to DM particles with 
cross-sections above a critical reference value \(\overline{\sigma}_{\rm ref.}^{e}\), which 
scales as \( \overline{\sigma}_{\rm ref.}^{e} \propto \epsilon^{2} \)~\cite{Emken:2019tni}. This reference critical 
cross section can be linked to bound on $\epsilon$ from below (for detail see e.~g.~Fig.~17 
from Ref.~\cite{Anchordoqui:2021ghd} and references therein).

Remarkably, the sufficiently small couplings, $\epsilon\to 0$, are constrained from above
due to DD bounds on the  scattering cross section. The latter can be 
seen from Fig.~10 in Ref.~\cite{Emken:2019tni} where the typical constraints on $\overline{\sigma}_e \propto \epsilon^2$ are shown from combined the analysis of above 
mentioned DD experiments. For clarity in Fig.~\ref{MesonNA64Limits}, the full excluded belt region is not displayed.

\section{Conclusions}\label{Conclusion}

In the present paper, by exploiting the state-of-the-art CalcHEP package we calculate 
the exact tree-level cross section of the lepton scattering off nucleus 
$l  N \to l  N \gamma^*(\to \chi \bar{\chi})$ to estimate the 
sensitivity of the NA64$\mu$ and LDMX fixed target experiments to MCP parameters. 
We show that   NA64$\mu$ experiment for  the expected statistics of muons incident on target 
$\mbox{MOT}\simeq  10^{14}$  can probe
the  MCP in the mass range  $10~\mbox{MeV} \lesssim m_\chi \lesssim 150~\mbox{MeV}$ with 
$10^{-4} \lesssim  \epsilon \lesssim 5\times 10^{-4}$ through the bremsstrahlung-like missing energy 
process $\mu N \to \mu N \gamma^{*}( \to \chi \bar{\chi})$.  We also estimated the expected sensitivity of the 
LDMX electron beam fixed target facility for the ultimate statistics of $\mbox{EOT}\simeq  2\times 10^{16}$ and 
electron beam of $E_{\rm e}\simeq 8~\mbox{GeV}$.  We demonstrated that for the LDMX 
 the relatively heavy MCP in the mass range $250~\mbox{MeV} \lesssim  m_\chi \lesssim  400 ~\mbox{MeV}$\,GeV and 
$10^{-3} \lesssim  \epsilon \lesssim  1.5 \times 10^{-3}$ can be probed from  the invisible $\rho$ meson 
decay signature  $e N \to e N \rho (\to \chi \bar{\chi})$. 
We emphasize that  probing millicharged particles with lepton missing momentum experiments requires
suppressing the background rate to a sufficiently small value, on the order of $\lesssim 10^{-16} - 10^{-14}$ per incident lepton,

The possible extension of the present work can be addressed to a probing of MCPs with NA64$\mu$ in the 
sub-MeV mass range~\cite{Arefyeva:2022eba}, implying that the stopping loss of the MCPs can impact these bounds. We leave that issue for future work.

\section{ Acknowledgements }  
We would like to thank S.~Demidov, R.~Dusaev, D.~Gorbunov, M.~Kirsanov, A.~Pukhov  for very helpful 
discussions and suggestions.   The work of DK and IV was supported by the Foundation for the
Advancement of Theoretical Physics and Mathematics
BASIS (Project No.~\text{24-1-2-11-2} and No.~\text{24-1-2-11-1}).
This research was supported by the Ministry of Education and Science of the Russian Federation in part of the Science program (Project FSWW-2023-0003). 
This work was funded by FONDECYT (Chile) under Grant 
No. 1240066 and by ANID$-$Millen\-nium Program$-$ICN2019\_044 (Chile).

\appendix

\section{Upper limit for number of signal events
\label{BayesianUpperLimit}}
This section outlines the procedure for determining the upper limit on the mean number of signal events, \( s \), in the presence of a known background, \( b \), using a Bayesian framework for Poisson statistics.

When the parameter of interest is constrained to be non-negative, \( s \geq 0 \), a conservative and common choice for 
the prior probability density,
\begin{equation}
\pi(s) = \theta(s) =
\begin{cases}
1, & s \geq 0 \\
0, & s < 0
\end{cases}
\label{PriorZeroBkg}
\end{equation}
is a uniform truncated distribution~\cite{ParticleDataGroup:2024cfk,Magill:2018tbb}. 
For cases where the signal is explicitly constrained to be greater than the background, \( s \gtrsim b \), an 
appropriate prior is the uniform distribution truncated below \( b \):

\begin{equation}
\pi(s) = \theta(s - b) =
\begin{cases}
1, & s \geq b \\
0, & s < b
\end{cases}
\label{Prior}
\end{equation}

Given this prior and an observed number of events, \( n_{\text{obs}} \), the likelihood function is given by the Poisson distribution with mean \( s + b \):

\begin{equation}
P(n_{\text{obs}} \mid s) = \frac{(s + b)^{n_{\text{obs}}}}{n_{\text{obs}}!} e^{-(s + b)}.
\label{PoissonLikelihood}
\end{equation}
The posterior probability density for \( s \) is proportional to the product of the likelihood (\ref{PoissonLikelihood}) and the prior (\ref{Prior}). The  upper limit, 
\( s_{\text{up}} \), at a Bayesian confidence level 
(in literature~\cite{ParticleDataGroup:2024cfk} it is called also a credibility level) of \( 1 - \alpha \) is then defined as the value 
for which the posterior probability that \( s \leq s_{\text{up}} \) is \( 1 - \alpha \). This is computed by solving:

\begin{equation}
1 - \alpha = \frac{\int\limits_{-\infty}^{s_{\text{up}}} P(n_{\text{obs}} \mid s) \, \pi(s) \, ds}{\int\limits_{-\infty}^{\infty} P(n_{\text{obs}} \mid s) \, \pi(s) \, ds}.
\label{eq:def_CL_limit}
\end{equation}

The integrals in Eq.~\eqref{eq:def_CL_limit} can be expressed in terms of the upper incomplete gamma function, 
$$ \Gamma(a, z)= \int\limits^\infty_z dt\, t^{a-1} e^{-t}   . 
$$
By evaluating the integrals with the truncated uniform prior \( \pi(s) = \theta(s - b) \), the upper limit \( s_{\text{up}} \) is extracted from the numerical solution to:

\begin{equation}
\alpha = \frac{\Gamma(n_{\text{obs}} + 1, \, s_{\text{up}} + b)}{\Gamma(n_{\text{obs}} + 1, \, 2b)}.
\label{IncomplGamma}
\end{equation}

\bibliography{bibl}

@techreport{NA64:2018iqr,
      author        = "Gninenko, Sergei",
      collaboration = "NA64 Collaboration",
      title         = "{Addendum to the Proposal P348: Search for dark sector particles weakly coupled to muon with NA64 \ensuremath{\mu}}",
      institution   = "CERN",
      reportNumber  = "CERN-SPSC-2018-024, SPSC-P-348-ADD-3",
      address       = "Geneva",
      year          = "2018",
      url           = "https://cds.cern.ch/record/2640930?ln=en"
}

@article{Allison:2025mom,
    author = "Allison, Elizabeth and Blinov, Nikita",
    title = "{Data-Driven Predictions for Dark Photon and Millicharged Particle Production}",
    eprint = "2512.04153",
    archivePrefix = "arXiv",
    primaryClass = "hep-ph",
    month = "12",
    year = "2025"
}

@article{DarkSide:2018ppu,
    author = "Agnes, P. and others",
    collaboration = "DarkSide",
    title = "{Constraints on Sub-GeV Dark-Matter{\textendash}Electron Scattering from the DarkSide-50 Experiment}",
    eprint = "1802.06998",
    archivePrefix = "arXiv",
    primaryClass = "astro-ph.CO",
    reportNumber = "FERMILAB-PUB-18-052-AD-AE-CD-E",
    doi = "10.1103/PhysRevLett.121.111303",
    journal = "Phys. Rev. Lett.",
    volume = "121",
    number = "11",
    pages = "111303",
    year = "2018"
}

@article{XENON:2016jmt,
    author = "Aprile, E. and others",
    collaboration = "XENON",
    title = "{Low-mass dark matter search using ionization signals in XENON100}",
    eprint = "1605.06262",
    archivePrefix = "arXiv",
    primaryClass = "astro-ph.CO",
    doi = "10.1103/PhysRevD.94.092001",
    journal = "Phys. Rev. D",
    volume = "94",
    number = "9",
    pages = "092001",
    year = "2016",
    note = "[Erratum: Phys.Rev.D 95, 059901 (2017)]"
}

@article{XENON10:2011prx,
    author = "Angle, J. and others",
    collaboration = "XENON10",
    title = "{A search for light dark matter in XENON10 data}",
    eprint = "1104.3088",
    archivePrefix = "arXiv",
    primaryClass = "astro-ph.CO",
    doi = "10.1103/PhysRevLett.107.051301",
    journal = "Phys. Rev. Lett.",
    volume = "107",
    pages = "051301",
    year = "2011",
    note = "[Erratum: Phys.Rev.Lett. 110, 249901 (2013)]"
}

@article{Essig:2017kqs,
    author = "Essig, Rouven and Volansky, Tomer and Yu, Tien-Tien",
    title = "{New Constraints and Prospects for sub-GeV Dark Matter Scattering off Electrons in Xenon}",
    eprint = "1703.00910",
    archivePrefix = "arXiv",
    primaryClass = "hep-ph",
    reportNumber = "CERN-TH-2017-042, YITP-SB-17-09",
    doi = "10.1103/PhysRevD.96.043017",
    journal = "Phys. Rev. D",
    volume = "96",
    number = "4",
    pages = "043017",
    year = "2017"
}

@article{SuperCDMS:2018mne,
    author = "Agnese, R. and others",
    collaboration = "SuperCDMS",
    title = "{First Dark Matter Constraints from a SuperCDMS Single-Charge Sensitive Detector}",
    eprint = "1804.10697",
    archivePrefix = "arXiv",
    primaryClass = "hep-ex",
    reportNumber = "FERMILAB-PUB-18-150-AE",
    doi = "10.1103/PhysRevLett.121.051301",
    journal = "Phys. Rev. Lett.",
    volume = "121",
    number = "5",
    pages = "051301",
    year = "2018",
    note = "[Erratum: Phys.Rev.Lett. 122, 069901 (2019)]"
}

@article{SENSEI:2019ibb,
    author = "Abramoff, Orr and others",
    collaboration = "SENSEI",
    title = "{SENSEI: Direct-Detection Constraints on Sub-GeV Dark Matter from a Shallow Underground Run Using a Prototype Skipper-CCD}",
    eprint = "1901.10478",
    archivePrefix = "arXiv",
    primaryClass = "hep-ex",
    reportNumber = "YITP-2019-01, FERMILAB-PUB-19-039-AE",
    doi = "10.1103/PhysRevLett.122.161801",
    journal = "Phys. Rev. Lett.",
    volume = "122",
    number = "16",
    pages = "161801",
    year = "2019"
}

@article{Crisler:2018gci,
    author = "Crisler, Michael and Essig, Rouven and Estrada, Juan and Fernandez, Guillermo and Tiffenberg, Javier and Sofo haro, Miguel and Volansky, Tomer and Yu, Tien-Tien",
    collaboration = "SENSEI",
    title = "{SENSEI: First Direct-Detection Constraints on sub-GeV Dark Matter from a Surface Run}",
    eprint = "1804.00088",
    archivePrefix = "arXiv",
    primaryClass = "hep-ex",
    reportNumber = "FERMILAB-PUB-18-116-AE-PPD, YITP-SB-18-4, CERN-TH-2018-070",
    doi = "10.1103/PhysRevLett.121.061803",
    journal = "Phys. Rev. Lett.",
    volume = "121",
    number = "6",
    pages = "061803",
    year = "2018"
}

@article{Bailloeul:2025fde,
    author = "Bailloeul, Leo and Citron, Matthew and Cui, Yanou and Foroughi-Abari, Saeid and Hwang, Insung and Li, Fengyi and Tsai, Yu-Dai and Liu, Ming Xiong and Gunthoti, Kranti and Yoo, Jae Hyeok",
    title = "{Dedicated Searches for Millicharged Particles at Intensity-Frontier Facilities: SpinQuest and SHiP}",
    eprint = "2512.11027",
    archivePrefix = "arXiv",
    primaryClass = "hep-ph",
    month = "12",
    year = "2025"
}

@article{Anchordoqui:2021ghd,
    author = "Anchordoqui, Luis A. and others",
    title = "{The Forward Physics Facility: Sites, experiments, and physics potential}",
    eprint = "2109.10905",
    archivePrefix = "arXiv",
    primaryClass = "hep-ph",
    reportNumber = "BNL-222142-2021-FORE, CERN-PBC-Notes-2021-025, DESY-21-142, DESY-21-142,
  FERMILAB-CONF-21-452-AE-E-ND-PPD-T, KYUSHU-RCAPP-2021-01, LU TP 21-36,
  PITT-PACC-2118, SMU-HEP-21-10, UCI-TR-2021-22, FERMILAB-CONF-21-452-AE-E-ND-PPD-T",
    doi = "10.1016/j.physrep.2022.04.004",
    journal = "Phys. Rept.",
    volume = "968",
    pages = "1--50",
    year = "2022"
}

@article{Kouvaris:2025tom,
    author = "Kouvaris, Chris and Shoemaker, Ian M.",
    title = "{Millicharged Particle Production in Pulsars via the Schwinger Effect}",
    eprint = "2511.04763",
    archivePrefix = "arXiv",
    primaryClass = "hep-ph",
    month = "11",
    year = "2025"
}

@article{Dmitrieva:2025ohn,
    author = "Dmitrieva, Ekaterina and Satunin, Petr",
    title = "{Resonant production of millicharged scalars in k{\textasciicircum}2 {\ensuremath{>}} 0 electromagnetic wave background}",
    eprint = "2510.25505",
    archivePrefix = "arXiv",
    primaryClass = "hep-ph",
    reportNumber = "INR-TH-2025-019",
    month = "10",
    year = "2025"
}

@article{Demidov:2025yyr,
    author = "Demidov, Sergey and Feschenko, Alexander and Gorbunov, Dmitry and Izmaylov, Alexander and Kalashnikov, Dmitry and Kravchuk, Leonid and Kriukova, Ekaterina and Kudenko, Yury and Mashin, Nikita and Senichev, Yury",
    title = "{Searches for new light particles at the Troitsk Meson Factory (TiMoFey)}",
    eprint = "2508.01968",
    archivePrefix = "arXiv",
    primaryClass = "hep-ph",
    reportNumber = "INR-TH-2025-010",
    month = "8",
    year = "2025"
}

@article{ParticleDataGroup:2024cfk,
    author = "Navas, S. and others",
    collaboration = "Particle Data Group",
    title = "{Review of particle physics}",
    doi = "10.1103/PhysRevD.110.030001",
    journal = "Phys. Rev. D",
    volume = "110",
    number = "3",
    pages = "030001",
    year = "2024"
}

@article{Brahm:1989jh,
    author = "Brahm, David E. and Hall, Lawrence J.",
    title = "{U(1)-prime DARK MATTER}",
    reportNumber = "LBL-27847, UCB-PTH-89/21",
    doi = "10.1103/PhysRevD.41.1067",
    journal = "Phys. Rev. D",
    volume = "41",
    pages = "1067",
    year = "1990"
}

@article{Chu:2018qrm,
    author = "Chu, Xiaoyong and Pradler, Josef and Semmelrock, Lukas",
    title = "{Light dark states with electromagnetic form factors}",
    eprint = "1811.04095",
    archivePrefix = "arXiv",
    primaryClass = "hep-ph",
    doi = "10.1103/PhysRevD.99.015040",
    journal = "Phys. Rev. D",
    volume = "99",
    number = "1",
    pages = "015040",
    year = "2019"
}

@article{LDMX:2025bog,
    author = "Akesson, Torsten and others",
    collaboration = "LDMX",
    title = "{LDMX - The Light Dark Matter eXperiment}",
    eprint = "2508.11833",
    archivePrefix = "arXiv",
    primaryClass = "hep-ex",
    reportNumber = "FERMILAB-PUB-25-0605-CSAID-PPD",
    month = "8",
    year = "2025"
}

@article{Berlin:2025hjs,
    author = "Berlin, Asher and Bogorad, Zachary and Graham, Peter W. and Ramani, Harikrishnan",
    title = "{Cavendish Tests of Millicharged Particles}",
    eprint = "2510.25825",
    archivePrefix = "arXiv",
    primaryClass = "hep-ph",
    reportNumber = "FERMILAB-PUB-25-0623-SQMS-T",
    month = "10",
    year = "2025"
}

@article{CONNIE:2024off,
    author = "Aguilar-Arevalo, Alexis A. and others",
    collaboration = "CONNIE, Atucha-II",
    title = "{Search for Reactor-Produced Millicharged Particles with Skipper-CCDs at the CONNIE and Atucha-II Experiments}",
    eprint = "2405.16316",
    archivePrefix = "arXiv",
    primaryClass = "hep-ex",
    reportNumber = "FERMILAB-PUB-24-0709-PPD",
    doi = "10.1103/PhysRevLett.134.071801",
    journal = "Phys. Rev. Lett.",
    volume = "134",
    number = "7",
    pages = "071801",
    year = "2025"
}

@article{Gao:2025ykc,
    author = "Gao, Ting and Pospelov, Maxim",
    title = "{Constraints on millicharged particles from nuclear gamma-decays}",
    eprint = "2507.17955",
    archivePrefix = "arXiv",
    primaryClass = "hep-ph",
    month = "7",
    year = "2025"
}

@article{Eberl:2025kfm,
    author = "Eberl, Helmut and Fahrecker, Maximilian and Pradler, Josef",
    title = "{Underground Production of Electromagnetic Dark States by MeV-scale Electron Beams and Detection with CCDs}",
    eprint = "2511.02023",
    archivePrefix = "arXiv",
    primaryClass = "hep-ph",
    reportNumber = "UWThPh 2025-20",
    month = "11",
    year = "2025"
}

@article{Berlin:2025btf,
    author = "Berlin, Asher and Bogorad, Zachary and Graham, Peter W. and Ramani, Harikrishnan",
    title = "{Electric Accumulation of Millicharged Particles}",
    eprint = "2510.25834",
    archivePrefix = "arXiv",
    primaryClass = "hep-ph",
    reportNumber = "FERMILAB-PUB-25-0622-SQMS-T",
    month = "10",
    year = "2025"
}

@article{Pospelov:2007mp,
    author = "Pospelov, Maxim and Ritz, Adam and Voloshin, Mikhail B.",
    title = "{Secluded WIMP Dark Matter}",
    eprint = "0711.4866",
    archivePrefix = "arXiv",
    primaryClass = "hep-ph",
    doi = "10.1016/j.physletb.2008.02.052",
    journal = "Phys. Lett. B",
    volume = "662",
    pages = "53--61",
    year = "2008"
}

@article{Feng:2022inv,
    author = "Feng, Jonathan L. and others",
    title = "{The Forward Physics Facility at the High-Luminosity LHC}",
    eprint = "2203.05090",
    archivePrefix = "arXiv",
    primaryClass = "hep-ex",
    reportNumber = "UCI-TR-2022-01, CERN-PBC-Notes-2022-001, INT-PUB-22-006, BONN-TH-2022-04, FERMILAB-PUB-22-094-ND-SCD-T",
    doi = "10.1088/1361-6471/ac865e",
    journal = "J. Phys. G",
    volume = "50",
    number = "3",
    pages = "030501",
    year = "2023"
}

@article{Tulin:2012wi,
    author = "Tulin, Sean and Yu, Hai-Bo and Zurek, Kathryn M.",
    title = "{Resonant Dark Forces and Small Scale Structure}",
    eprint = "1210.0900",
    archivePrefix = "arXiv",
    primaryClass = "hep-ph",
    reportNumber = "MCTP-12-27",
    doi = "10.1103/PhysRevLett.110.111301",
    journal = "Phys. Rev. Lett.",
    volume = "110",
    number = "11",
    pages = "111301",
    year = "2013"
}

@article{Feng:2009mn,
    author = "Feng, Jonathan L. and Kaplinghat, Manoj and Tu, Huitzu and Yu, Hai-Bo",
    title = "{Hidden Charged Dark Matter}",
    eprint = "0905.3039",
    archivePrefix = "arXiv",
    primaryClass = "hep-ph",
    reportNumber = "UCI-TR-2009-06",
    doi = "10.1088/1475-7516/2009/07/004",
    journal = "JCAP",
    volume = "07",
    pages = "004",
    year = "2009"
}

@article{Harnik:2019zee,
    author = "Harnik, Roni and Liu, Zhen and Palamara, Ornella",
    title = "{Millicharged Particles in Liquid Argon Neutrino Experiments}",
    eprint = "1902.03246",
    archivePrefix = "arXiv",
    primaryClass = "hep-ph",
    reportNumber = "FERMILAB-PUB-19-060-ND-T",
    doi = "10.1007/JHEP07(2019)170",
    journal = "JHEP",
    volume = "07",
    pages = "170",
    year = "2019"
}

@article{Holdom:1985ag,
    author = "Holdom, Bob",
    title = "{Two U(1)'s and Epsilon Charge Shifts}",
    reportNumber = "UTPT-85-30",
    doi = "10.1016/0370-2693(86)91377-8",
    journal = "Phys. Lett. B",
    volume = "166",
    pages = "196--198",
    year = "1986"
}

@article{Feldman:2007wj,
    author = "Feldman, Daniel and Liu, Zuowei and Nath, Pran",
    title = "{The Stueckelberg Z-prime Extension with Kinetic Mixing and Milli-Charged Dark Matter From the Hidden Sector}",
    eprint = "hep-ph/0702123",
    archivePrefix = "arXiv",
    doi = "10.1103/PhysRevD.75.115001",
    journal = "Phys. Rev. D",
    volume = "75",
    pages = "115001",
    year = "2007"
}

@article{Cline:2012is,
    author = "Cline, James M. and Liu, Zuowei and Xue, Wei",
    title = "{Millicharged Atomic Dark Matter}",
    eprint = "1201.4858",
    archivePrefix = "arXiv",
    primaryClass = "hep-ph",
    doi = "10.1103/PhysRevD.85.101302",
    journal = "Phys. Rev. D",
    volume = "85",
    pages = "101302",
    year = "2012"
}

@article{SENSEI:2023gie,
    author = "Barak, Liron and others",
    collaboration = "SENSEI",
    title = "{Search by the SENSEI Experiment for Millicharged Particles Produced in the NuMI Beam}",
    eprint = "2305.04964",
    archivePrefix = "arXiv",
    primaryClass = "hep-ex",
    reportNumber = "CALT-TH-2023-011, YITP-SB-2023-07, FERMILAB-PUB-23-222-PPD",
    doi = "10.1103/PhysRevLett.133.071801",
    journal = "Phys. Rev. Lett.",
    volume = "133",
    number = "7",
    pages = "071801",
    year = "2024"
}

@article{Liu:2018jdi,
    author = "Liu, Zuowei and Zhang, Yu",
    title = "{Probing millicharge at BESIII via monophoton searches}",
    eprint = "1808.00983",
    archivePrefix = "arXiv",
    primaryClass = "hep-ph",
    doi = "10.1103/PhysRevD.99.015004",
    journal = "Phys. Rev. D",
    volume = "99",
    number = "1",
    pages = "015004",
    year = "2019"
}

@article{Zhang:2019wnz,
    author = "Zhang, Yu and Zhang, Wei-Tao and Song, Mao and Pan, Xue-An and Niu, Zhong-Ming and Li, Gang",
    title = "{Probing invisible decay of dark photon at BESIII and future STCF via monophoton searches}",
    eprint = "1907.07046",
    archivePrefix = "arXiv",
    primaryClass = "hep-ph",
    doi = "10.1103/PhysRevD.100.115016",
    journal = "Phys. Rev. D",
    volume = "100",
    number = "11",
    pages = "115016",
    year = "2019"
}

@article{Liu:2019ogn,
    author = "Liu, Zuowei and Xu, Yong-Heng and Zhang, Yu",
    title = "{Probing dark matter particles at CEPC}",
    eprint = "1903.12114",
    archivePrefix = "arXiv",
    primaryClass = "hep-ph",
    doi = "10.1007/JHEP06(2019)009",
    journal = "JHEP",
    volume = "06",
    pages = "009",
    year = "2019"
}

@article{Liang:2019zkb,
    author = "Liang, Jinhan and Liu, Zuowei and Ma, Yue and Zhang, Yu",
    title = "{Millicharged particles at electron colliders}",
    eprint = "1909.06847",
    archivePrefix = "arXiv",
    primaryClass = "hep-ph",
    doi = "10.1103/PhysRevD.102.015002",
    journal = "Phys. Rev. D",
    volume = "102",
    number = "1",
    pages = "015002",
    year = "2020"
}

@article{Bai:2021nai,
    author = "Bai, Yang and Lee, Seung J. and Son, Minho and Ye, Fang",
    title = "{Muon g \ensuremath{-} 2 from millicharged hidden confining sector}",
    eprint = "2106.15626",
    archivePrefix = "arXiv",
    primaryClass = "hep-ph",
    doi = "10.1007/JHEP11(2021)019",
    journal = "JHEP",
    volume = "11",
    pages = "019",
    year = "2021"
}

@article{milliQan:2021lne,
    author = "Ball, A. and others",
    collaboration = "milliQan",
    title = "{Sensitivity to millicharged particles in future proton-proton collisions at the LHC with the milliQan detector}",
    eprint = "2104.07151",
    archivePrefix = "arXiv",
    primaryClass = "hep-ex",
    doi = "10.1103/PhysRevD.104.032002",
    journal = "Phys. Rev. D",
    volume = "104",
    number = "3",
    pages = "032002",
    year = "2021"
}

@article{Arefyeva:2022eba,
    author = "Arefyeva, Nataliya and Gninenko, Sergei and Gorbunov, Dmitry and Kirpichnikov, Dmitry",
    title = "{Passage of millicharged particles in the electron beam-dump: Refining constraints from SLACmQ and estimating sensitivity of NA64e}",
    eprint = "2204.03984",
    archivePrefix = "arXiv",
    primaryClass = "hep-ph",
    reportNumber = "INR-TH-2022-008",
    doi = "10.1103/PhysRevD.106.035029",
    journal = "Phys. Rev. D",
    volume = "106",
    number = "3",
    pages = "035029",
    year = "2022"
}

@article{Chu:2020ysb,
    author = "Chu, Xiaoyong and Kuo, Jui-Lin and Pradler, Josef",
    title = "{Dark sector-photon interactions in proton-beam experiments}",
    eprint = "2001.06042",
    archivePrefix = "arXiv",
    primaryClass = "hep-ph",
    doi = "10.1103/PhysRevD.101.075035",
    journal = "Phys. Rev. D",
    volume = "101",
    number = "7",
    pages = "075035",
    year = "2020"
}

@article{Gorbunov:2021jog,
    author = "Gorbunov, Dmitry and Krasnov, Igor and Kudenko, Yury and Suvorov, Sergey",
    title = "{Double-hit signature of millicharged particles in 3D segmented neutrino detector}",
    eprint = "2103.11814",
    archivePrefix = "arXiv",
    primaryClass = "hep-ph",
    reportNumber = "INR-TH-2021-006",
    doi = "10.1016/j.physletb.2021.136641",
    journal = "Phys. Lett. B",
    volume = "822",
    pages = "136641",
    year = "2021"
}

@article{ArgoNeuT:2019ckq,
    author = "Acciarri, R. and others",
    collaboration = "ArgoNeuT",
    title = "{Improved Limits on Millicharged Particles Using the ArgoNeuT Experiment at Fermilab}",
    eprint = "1911.07996",
    archivePrefix = "arXiv",
    primaryClass = "hep-ex",
    reportNumber = "FERMILAB-PUB-19-582-ND",
    doi = "10.1103/PhysRevLett.124.131801",
    journal = "Phys. Rev. Lett.",
    volume = "124",
    number = "13",
    pages = "131801",
    year = "2020"
}

@article{Harnik:2020ugb,
    author = "Harnik, Roni and Plestid, Ryan and Pospelov, Maxim and Ramani, Harikrishnan",
    title = "{Millicharged cosmic rays and low recoil detectors}",
    eprint = "2010.11190",
    archivePrefix = "arXiv",
    primaryClass = "hep-ph",
    reportNumber = "FERMILAB-PUB-20-523-T",
    doi = "10.1103/PhysRevD.103.075029",
    journal = "Phys. Rev. D",
    volume = "103",
    number = "7",
    pages = "075029",
    year = "2021"
}

@article{Aboubrahim:2021ohe,
    author = "Aboubrahim, Amin and Nath, Pran and Wang, Zhu-Yao",
    title = "{A cosmologically consistent millicharged dark matter solution to the EDGES anomaly of possible string theory origin}",
    eprint = "2108.05819",
    archivePrefix = "arXiv",
    primaryClass = "hep-ph",
    reportNumber = "MS-TP-21-27",
    doi = "10.1007/JHEP12(2021)148",
    journal = "JHEP",
    volume = "12",
    pages = "148",
    year = "2021"
}

@article{Li:2020wyl,
    author = "Li, Jung-Tsung and Lin, Tongyan",
    title = "{Dynamics of millicharged dark matter in supernova remnants}",
    eprint = "2002.04625",
    archivePrefix = "arXiv",
    primaryClass = "astro-ph.CO",
    doi = "10.1103/PhysRevD.101.103034",
    journal = "Phys. Rev. D",
    volume = "101",
    number = "10",
    pages = "103034",
    year = "2020"
}

@article{Caputo:2019tms,
    author = "Caputo, Andrea and Sberna, Laura and Frias, Miguel and Blas, Diego and Pani, Paolo and Shao, Lijing and Yan, Wenming",
    title = "{Constraints on millicharged dark matter and axionlike particles from timing of radio waves}",
    eprint = "1902.02695",
    archivePrefix = "arXiv",
    primaryClass = "astro-ph.CO",
    doi = "10.1103/PhysRevD.100.063515",
    journal = "Phys. Rev. D",
    volume = "100",
    number = "6",
    pages = "063515",
    year = "2019"
}

@article{ArguellesDelgado:2021lek,
    author = {Arg\"uelles Delgado, Carlos Alberto and Kelly, Kevin James and Mu\~noz Albornoz, V\'\i{}ctor},
    title = "{Millicharged particles from the heavens: single- and multiple-scattering signatures}",
    eprint = "2104.13924",
    archivePrefix = "arXiv",
    primaryClass = "hep-ph",
    reportNumber = "FERMILAB-PUB-21-214-T",
    doi = "10.1007/JHEP11(2021)099",
    journal = "JHEP",
    volume = "11",
    pages = "099",
    year = "2021"
}

@article{TEXONO:2018nir,
    author = "Singh, L. and others",
    collaboration = "TEXONO",
    title = "{Constraints on millicharged particles with low threshold germanium detectors at Kuo-Sheng Reactor Neutrino Laboratory}",
    eprint = "1808.02719",
    archivePrefix = "arXiv",
    primaryClass = "hep-ph",
    doi = "10.1103/PhysRevD.99.032009",
    journal = "Phys. Rev. D",
    volume = "99",
    number = "3",
    pages = "032009",
    year = "2019"
}

@article{Chen:2014dsa,
    author = "Chen, Jiunn-Wei and Chi, Hsin-Chang and Li, Hau-Bin and Liu, C. -P. and Singh, Lakhwinder and Wong, Henry T. and Wu, Chih-Liang and Wu, Chih-Pan",
    title = "{Constraints on millicharged neutrinos via analysis of data from atomic ionizations with germanium detectors at sub-keV sensitivities}",
    eprint = "1405.7168",
    archivePrefix = "arXiv",
    primaryClass = "hep-ph",
    doi = "10.1103/PhysRevD.90.011301",
    journal = "Phys. Rev. D",
    volume = "90",
    number = "1",
    pages = "011301",
    year = "2014"
}

@article{MiniBooNE:2018esg,
    author = "Aguilar-Arevalo, A. A. and others",
    collaboration = "MiniBooNE",
    title = "{Significant Excess of ElectronLike Events in the MiniBooNE Short-Baseline Neutrino Experiment}",
    eprint = "1805.12028",
    archivePrefix = "arXiv",
    primaryClass = "hep-ex",
    reportNumber = "LA-UR-18-24586, FERMILAB-PUB-18-219-AD-PPD-ND",
    doi = "10.1103/PhysRevLett.121.221801",
    journal = "Phys. Rev. Lett.",
    volume = "121",
    number = "22",
    pages = "221801",
    year = "2018"
}

@article{MicroBooNE:2017kvv,
    author = "Acciarri, R. and others",
    collaboration = "MicroBooNE",
    title = "{Michel Electron Reconstruction Using Cosmic-Ray Data from the MicroBooNE LArTPC}",
    eprint = "1704.02927",
    archivePrefix = "arXiv",
    primaryClass = "physics.ins-det",
    reportNumber = "FERMILAB-PUB-17-090-ND",
    doi = "10.1088/1748-0221/12/09/P09014",
    journal = "JINST",
    volume = "12",
    number = "09",
    pages = "P09014",
    year = "2017"
}

@article{Magill:2018tbb,
    author = "Magill, Gabriel and Plestid, Ryan and Pospelov, Maxim and Tsai, Yu-Dai",
    title = "{Millicharged particles in neutrino experiments}",
    eprint = "1806.03310",
    archivePrefix = "arXiv",
    primaryClass = "hep-ph",
    reportNumber = "FERMILAB-PUB-18-631-A",
    doi = "10.1103/PhysRevLett.122.071801",
    journal = "Phys. Rev. Lett.",
    volume = "122",
    number = "7",
    pages = "071801",
    year = "2019"
}

@article{Plestid:2020kdm,
    author = "Plestid, Ryan and Takhistov, Volodymyr and Tsai, Yu-Dai and Bringmann, Torsten and Kusenko, Alexander and Pospelov, Maxim",
    title = "{New Constraints on Millicharged Particles from Cosmic-ray Production}",
    eprint = "2002.11732",
    archivePrefix = "arXiv",
    primaryClass = "hep-ph",
    reportNumber = "FERMILAB-PUB-20-044-A-T, INT-PUB-20-004, IPMU20-0015",
    doi = "10.1103/PhysRevD.102.115032",
    journal = "Phys. Rev. D",
    volume = "102",
    pages = "115032",
    year = "2020"
}

@article{Marocco:2020dqu,
    author = "Marocco, Giacomo and Sarkar, Subir",
    title = "{Blast from the past: Constraints on the dark sector from the BEBC WA66 beam dump experiment}",
    eprint = "2011.08153",
    archivePrefix = "arXiv",
    primaryClass = "hep-ph",
    doi = "10.21468/SciPostPhys.10.2.043",
    journal = "SciPost Phys.",
    volume = "10",
    number = "2",
    pages = "043",
    year = "2021"
}

@article{Prinz:1998ua,
    author = "Prinz, A. A. and others",
    title = "{Search for millicharged particles at SLAC}",
    eprint = "hep-ex/9804008",
    archivePrefix = "arXiv",
    reportNumber = "SLAC-PUB-7762",
    doi = "10.1103/PhysRevLett.81.1175",
    journal = "Phys. Rev. Lett.",
    volume = "81",
    pages = "1175--1178",
    year = "1998"
}

@article{LSND:2001akn,
    author = "Auerbach, L. B. and others",
    collaboration = "LSND",
    title = "{Measurement of electron - neutrino - electron elastic scattering}",
    eprint = "hep-ex/0101039",
    archivePrefix = "arXiv",
    doi = "10.1103/PhysRevD.63.112001",
    journal = "Phys. Rev. D",
    volume = "63",
    pages = "112001",
    year = "2001"
}

@article{Gninenko:2018ter,
    author = "Gninenko, S. N. and Kirpichnikov, D. V. and Krasnikov, N. V.",
    title = "{Probing millicharged particles with NA64 experiment at CERN}",
    eprint = "1810.06856",
    archivePrefix = "arXiv",
    primaryClass = "hep-ph",
    doi = "10.1103/PhysRevD.100.035003",
    journal = "Phys. Rev. D",
    volume = "100",
    number = "3",
    pages = "035003",
    year = "2019"
}

@article{Kim:1973he,
    author = "Kim, Kwang Je and Tsai, Yung-Su",
    title = "{IMPROVED WEIZSACKER-WILLIAMS METHOD AND ITS APPLICATION TO LEPTON AND W BOSON PAIR PRODUCTION}",
    reportNumber = "SLAC-PUB-1106-rev, SLAC-PUB-1106",
    doi = "10.1103/PhysRevD.8.3109",
    journal = "Phys. Rev. D",
    volume = "8",
    pages = "3109",
    year = "1973"
}

@article{Belyaev:2012qa,
    author = "Belyaev, Alexander and Christensen, Neil D. and Pukhov, Alexander",
    title = "{CalcHEP 3.4 for collider physics within and beyond the Standard Model}",
    eprint = "1207.6082",
    archivePrefix = "arXiv",
    primaryClass = "hep-ph",
    reportNumber = "PITT-PACC-1209",
    doi = "10.1016/j.cpc.2013.01.014",
    journal = "Comput. Phys. Commun.",
    volume = "184",
    pages = "1729--1769",
    year = "2013"
}

@article{Sieber:2021fue,
    author = "Sieber, H. and Banerjee, D. and Crivelli, P. and Depero, E. and Gninenko, S. N. and Kirpichnikov, D. V. and Kirsanov, M. M. and Poliakov, V. and Molina Bueno, L.",
    title = "{Prospects in the search for a new light Z' boson with the NA64\ensuremath{\mu} experiment at the CERN SPS}",
    eprint = "2110.15111",
    archivePrefix = "arXiv",
    primaryClass = "hep-ex",
    doi = "10.1103/PhysRevD.105.052006",
    journal = "Phys. Rev. D",
    volume = "105",
    number = "5",
    pages = "052006",
    year = "2022"
}

@article{Essig:2024dpa,
    author = "Essig, Rouven and Li, Peiran and Liu, Zhen and McDuffie, Megan and Plestid, Ryan and Xu, Hailin",
    title = "{Probing millicharged particles at an electron beam dump with ultralow-threshold sensors}",
    eprint = "2412.09652",
    archivePrefix = "arXiv",
    primaryClass = "hep-ph",
    doi = "10.1007/JHEP04(2025)057",
    journal = "JHEP",
    volume = "04",
    pages = "057",
    year = "2025"
}

@article{Forbes:2024zks,
    author = "Forbes, Diana and Kahn, Yonatan and Nguyen, Rachel",
    title = "{Exotic particles at the DUNE near detector from charged pion scattering}",
    eprint = "2407.14648",
    archivePrefix = "arXiv",
    primaryClass = "hep-ph",
    doi = "10.1103/PhysRevD.110.095029",
    journal = "Phys. Rev. D",
    volume = "110",
    number = "9",
    pages = "095029",
    year = "2024"
}

@article{Liu:2019knx,
    author = "Liu, Hongwan and Outmezguine, Nadav Joseph and Redigolo, Diego and Volansky, Tomer",
    title = "{Reviving Millicharged Dark Matter for 21-cm Cosmology}",
    eprint = "1908.06986",
    archivePrefix = "arXiv",
    primaryClass = "hep-ph",
    reportNumber = "MIT-CTP/5126",
    doi = "10.1103/PhysRevD.100.123011",
    journal = "Phys. Rev. D",
    volume = "100",
    number = "12",
    pages = "123011",
    year = "2019"
}

@article{Emken:2019tni,
    author = "Emken, Timon and Essig, Rouven and Kouvaris, Chris and Sholapurkar, Mukul",
    title = "{Direct Detection of Strongly Interacting Sub-GeV Dark Matter via Electron Recoils}",
    eprint = "1905.06348",
    archivePrefix = "arXiv",
    primaryClass = "hep-ph",
    reportNumber = "CERN-TH-2019-071, CP3-Origins-2019-18 DNRF90, YITP-SB-19-14",
    doi = "10.1088/1475-7516/2019/09/070",
    journal = "JCAP",
    volume = "09",
    pages = "070",
    year = "2019"
}

@article{Gninenko:2014pea,
    author = "Gninenko, S. N. and Krasnikov, N. V. and Matveev, V. A.",
    title = "{Muon g-2 and searches for a new leptophobic sub-GeV dark boson in a missing-energy experiment at CERN}",
    eprint = "1412.1400",
    archivePrefix = "arXiv",
    primaryClass = "hep-ph",
    doi = "10.1103/PhysRevD.91.095015",
    journal = "Phys. Rev. D",
    volume = "91",
    pages = "095015",
    year = "2015"
}

@article{Gninenko:2018tlp,
    author = "Gninenko, S. N. and Krasnikov, N. V.",
    title = "{Probing the muon $g_\mu$ - 2 anomaly, $L_\mu  -  L_\tau$ gauge boson and Dark Matter in dark photon experiments}",
    eprint = "1801.10448",
    archivePrefix = "arXiv",
    primaryClass = "hep-ph",
    doi = "10.1016/j.physletb.2018.06.043",
    journal = "Phys. Lett. B",
    volume = "783",
    pages = "24--28",
    year = "2018"
}

@article{Kirpichnikov:2021jev,
    author = "Kirpichnikov, D. V. and Sieber, H. and Bueno, L. Molina and Crivelli, P. and Kirsanov, M. M.",
    title = "{Probing hidden sectors with a muon beam: Total and differential cross sections for vector boson production in muon bremsstrahlung}",
    eprint = "2107.13297",
    archivePrefix = "arXiv",
    primaryClass = "hep-ph",
    doi = "10.1103/PhysRevD.104.076012",
    journal = "Phys. Rev. D",
    volume = "104",
    number = "7",
    pages = "076012",
    year = "2021"
}

@article{Voronchikhin:2024vfu,
    author = "Voronchikhin, I. V. and Kirpichnikov, D. V.",
    title = "{Implication of the Weizsacker-Williams approximation for the dark matter mediator production}",
    eprint = "2409.12748",
    archivePrefix = "arXiv",
    primaryClass = "hep-ph",
    doi = "10.1103/PhysRevD.111.035034",
    journal = "Phys. Rev. D",
    volume = "111",
    number = "3",
    pages = "035034",
    year = "2025"
}

@article{Mans:2017vej,
  author        = "M{\o}lgaard, Andreas and others",
  title         = "{Light Dark Matter eXperiment (LDMX)}",
  collaboration = "LDMX",
  year          = "2017",
  eprint        = "1709.01594",
  archivePrefix = "arXiv",
  primaryClass  = "hep-ex",
  reportNumber  = "FERMILAB-CONF-17-468-PPD"
}

@article{Berlin:2018bsc,
  author        = "Berlin, Asher and others",
  title         = "{Dark Matter, Millicharges, Axion and Scalar Particles, Gauge Bosons, and Other New Physics with LDMX}",
  journal       = "Phys. Rev. D",
  volume        = "99",
  year          = "2019",
  pages         = "075001",
  doi           = "10.1103/PhysRevD.99.075001",
  eprint        = "1807.01730",
  archivePrefix = "arXiv",
  primaryClass  = "hep-ph"
}

@inproceedings{Akesson:2022vza,
    author = "\r{A}kesson, Torsten and others",
    title = "{Current Status and Future Prospects for the Light Dark Matter eXperiment}",
    booktitle = "{Snowmass 2021}",
    eprint = "2203.08192",
    archivePrefix = "arXiv",
    primaryClass = "hep-ex",
    reportNumber = "FERMILAB-CONF-22-313-PPD-SCD-T",
    month = "3",
    year = "2022"
}

@article{LDMX:2018cma,
  collaboration = "LDMX",
  title         = "{Light Dark Matter eXperiment (LDMX): Technical Design Report}",
  year          = "2018",
  eprint        = "1808.05219",
  archivePrefix = "arXiv",
  primaryClass  = "physics.ins-det",
  reportNumber  = "FERMILAB-TM-2736-AE-CD"
}

@article{Ankowski:2019mfd,
    author = "Ankowski, Artur M. and Friedland, Alexander and Li, Shirley Weishi and Moreno, Omar and Schuster, Philip and Toro, Natalia and Tran, Nhan",
    title = "{Lepton-Nucleus Cross Section Measurements for DUNE with the LDMX Detector}",
    eprint = "1912.06140",
    archivePrefix = "arXiv",
    primaryClass = "hep-ph",
    reportNumber = "SLAC-PUB-17494, FERMILAB-PUB-19-619-SCD",
    doi = "10.1103/PhysRevD.101.053004",
    journal = "Phys. Rev. D",
    volume = "101",
    number = "5",
    pages = "053004",
    year = "2020"
}

@article{Schuster:2021mlr,
    author = "Schuster, Philip and Toro, Natalia and Zhou, Kevin",
    title = "{Probing invisible vector meson decays with the NA64 and LDMX experiments}",
    eprint = "2112.02104",
    archivePrefix = "arXiv",
    primaryClass = "hep-ph",
    reportNumber = "SLAC-PUB-17635",
    doi = "10.1103/PhysRevD.105.035036",
    journal = "Phys. Rev. D",
    volume = "105",
    number = "3",
    pages = "035036",
    year = "2022"
}

@article{NA64:2024nwj,
    author = "Andreev, Yu. M. and others",
    collaboration = "NA64",
    title = "{Shedding light on dark sectors with high-energy muons at the NA64 experiment at the CERN SPS}",
    eprint = "2409.10128",
    archivePrefix = "arXiv",
    primaryClass = "hep-ex",
    reportNumber = "CERN-EP-2024-236",
    doi = "10.1103/PhysRevD.110.112015",
    journal = "Phys. Rev. D",
    volume = "110",
    number = "11",
    pages = "112015",
    year = "2024"
}

@article{NA64:2024klw,
    author = "Andreev, Yu. M. and others",
    collaboration = "NA64",
    title = "{First Results in the Search for Dark Sectors at NA64 with the CERN SPS High Energy Muon Beam}",
    eprint = "2401.01708",
    archivePrefix = "arXiv",
    primaryClass = "hep-ex",
    reportNumber = "CERN-EP-2023-306",
    doi = "10.1103/PhysRevLett.132.211803",
    journal = "Phys. Rev. Lett.",
    volume = "132",
    number = "21",
    pages = "211803",
    year = "2024"
}

@article{Gninenko:2017yus,
    author = "Gninenko, S. N. and Kirpichnikov, D. V. and Kirsanov, M. M. and Krasnikov, N. V.",
    title = "{The exact tree-level calculation of the dark photon production in high-energy electron scattering at the CERN SPS}",
    eprint = "1712.05706",
    archivePrefix = "arXiv",
    primaryClass = "hep-ph",
    doi = "10.1016/j.physletb.2018.05.010",
    journal = "Phys. Lett. B",
    volume = "782",
    pages = "406--411",
    year = "2018"
}

@article{Gninenko:2019qiv,
    author = "Gninenko, S. N. and Kirpichnikov, D. V. and Kirsanov, M. M. and Krasnikov, N. V.",
    title = "{Combined search for light dark matter with electron and muon beams at NA64}",
    eprint = "1903.07899",
    archivePrefix = "arXiv",
    primaryClass = "hep-ph",
    doi = "10.1016/j.physletb.2019.07.015",
    journal = "Phys. Lett. B",
    volume = "796",
    pages = "117--122",
    year = "2019"
}

@article{Banerjee:2019pds,
    author = "Banerjee, D. and others",
    title = "{Dark matter search in missing energy events with NA64}",
    eprint = "1906.00176",
    archivePrefix = "arXiv",
    primaryClass = "hep-ex",
    reportNumber = "CERN-EP-2019-116",
    doi = "10.1103/PhysRevLett.123.121801",
    journal = "Phys. Rev. Lett.",
    volume = "123",
    number = "12",
    pages = "121801",
    year = "2019"
}

@article{NA64:2021xzo,
    author = "Andreev, Yu. M. and others",
    collaboration = "NA64",
    title = "{Constraints on New Physics in Electron $g-2$ from a Search for Invisible Decays of a Scalar, Pseudoscalar, Vector, and Axial Vector}",
    eprint = "2102.01885",
    archivePrefix = "arXiv",
    primaryClass = "hep-ex",
    reportNumber = "CERN-EP-2021-017",
    doi = "10.1103/PhysRevLett.126.211802",
    journal = "Phys. Rev. Lett.",
    volume = "126",
    number = "21",
    pages = "211802",
    year = "2021"
}

@article{Andreev:2021fzd,
    author = "Andreev, Yu. M. and others",
    title = "{Improved exclusion limit for light dark matter from e+e- annihilation in NA64}",
    eprint = "2108.04195",
    archivePrefix = "arXiv",
    primaryClass = "hep-ex",
    reportNumber = "CERN-EP-2021-164",
    doi = "10.1103/PhysRevD.104.L091701",
    journal = "Phys. Rev. D",
    volume = "104",
    number = "9",
    pages = "L091701",
    year = "2021"
}

@article{NA64:2023wbi,
    author = "Andreev, Yu. M. and others",
    collaboration = "NA64",
    title = "{Search for Light Dark Matter with NA64 at CERN}",
    eprint = "2307.02404",
    archivePrefix = "arXiv",
    primaryClass = "hep-ex",
    reportNumber = "CERN-EP-2023-130",
    doi = "10.1103/PhysRevLett.131.161801",
    journal = "Phys. Rev. Lett.",
    volume = "131",
    number = "16",
    pages = "161801",
    year = "2023"
}

@article{NA64:2022yly,
    author = "Andreev, Yu. M. and others",
    collaboration = "NA64",
    title = "{Search for a New B-L Z' Gauge Boson with the NA64 Experiment at CERN}",
    eprint = "2207.09979",
    archivePrefix = "arXiv",
    primaryClass = "hep-ex",
    reportNumber = "CERN-EP-2022-156",
    doi = "10.1103/PhysRevLett.129.161801",
    journal = "Phys. Rev. Lett.",
    volume = "129",
    number = "16",
    pages = "161801",
    year = "2022"
}

@article{Chen:2017awl,
    author = "Chen, Chien-Yi and Pospelov, Maxim and Zhong, Yi-Ming",
    title = "{Muon Beam Experiments to Probe the Dark Sector}",
    eprint = "1701.07437",
    archivePrefix = "arXiv",
    primaryClass = "hep-ph",
    doi = "10.1103/PhysRevD.95.115005",
    journal = "Phys. Rev. D",
    volume = "95",
    number = "11",
    pages = "115005",
    year = "2017"
}

\end{document}